\documentclass{qjrms2}
\usepackage{colordvi,graphicx,latexsym}
\allowdisplaybreaks

\def\MM#1{\boldsymbol{#1}}

\begin{document}

\QJRMS{1}{999}{128}{2006}{yy.n}

\runningheads{C. J. Cotter \and\ J. Frank 
\and\ S. Reich}{The remapped particle-mesh 
advection scheme}

\title{The remapped particle-mesh semi-Lagrangian advection scheme}

\author{C. J. Cotter$^1$ and J. Frank$^2$  and 
S. Reich$^3$\footnote{Corresponding author: Universit\"at Potsdam, 
Institut f\"ur Mathematik, Postfach 60 15 53, D-14415 Potsdam, 
Germany (e-mail: \texttt{sreich@math.uni-potsdam.de})}}

\affiliation{$^1$Imperial College London, United Kingdom\\
$^2$CWI Amsterdam, The Netherlands\\
$^3$Universit\"at Potsdam, Germany} 

\date{(Received 1 January 0000; revised 31 January 0001)}

\rmscopyright 

\begin{abstract}
We describe the \emph{remapped particle-mesh} method, a new
mass-conserving method for solving the density equation which is
suitable for combining with semi-Lagrangian methods for compressible
flow applied to numerical weather prediction. In addition to the
conservation property, the remapped particle-mesh method is computationally
efficient and at least as accurate as current semi-Lagrangian methods based on
cubic interpolation. We provide results of tests of the method in the
plane, results from incorporating the advection method into a
semi-Lagrangian method for the rotating shallow-water equations
in planar geometry, and results from extending the method to 
the surface of a sphere.
\end{abstract}


\tableofcontents

\ahead{Introduction}

The \emph{semi-implicit semi-Lagrangian} (SISL) method, as originally
introduced by {\sc Robert} \cite{Robert82}, has become very popular
in numerical weather prediction (NWP). The semi-Lagrangian aspect
of SISL schemes allows for a relatively accurate treatment of advection
while at the same time avoiding step size restrictions of explicit
Eulerian methods. The standard semi-Lagrangian algorithm 
(see, e.g., \cite{Staniforth91}) 
calculates departure points, i.e., the positions of Lagrangian 
particles which will be advected onto the grid during the time step. 
The momentum and density equations are then solved along the 
trajectory of the particles. This calculation requires interpolation 
to obtain velocity and density values at the departure point.  It 
has been found that cubic interpolation is both accurate and 
computationally tractable (see, e.g., \cite{Staniforth91}).

Ideally, as well as being efficient and accurate, a
density advection scheme should exactly preserve mass in order to be
useful for, e.g., climate prediction or atmospheric chemistry
calculations. Recent developments have involved computing the change
in volume elements, defined between departure and arrival points, 
making use of a technique called cascade interpolation 
\cite{Purser}. Several such methods have been suggested 
in recent years, including the methods of {\sc Nair et al} 
\cite{Nair1,Nair2,Nair3} and the SLICE schemes
of {\sc Zerroukat et al} \cite{Zerroukat1,Zerroukat2,Zerroukat3,Zerroukat4}.

In this paper we give a new density advection scheme, the remapped
particle-mesh method, which is based on the particle-mesh
discretisation for the density equation used in the Hamiltonian
Particle-Mesh (HPM) method suggested by {\sc Gottwald, Frank \& Reich} 
\cite{frank02sp}, which itself was a
combination of smoothed particle-hydrodynamics
\cite{lucy77,gingold77} and particle-in-cell methods \cite{har64}. 
The particle-mesh method provides a very simple discretisation which
conserves mass by construction, and may be adapted to nonplanar
geometries such as the sphere \cite{FrRe03}. In this paper we show
that an efficient scheme can be obtained by mapping the
particles back to the grid after each time step. Our numerical
results show that this scheme is at least 
as accurate as standard semi-Lagrangian advection 
using cubic interpolation at departure points. We show how the method may be 
included in the staggered semi-Lagrangian schemes, 
proposed by {\sc Staniforth et al} \cite{ASL4} and {\sc Reich} 
\cite{reich_bit06}, and show how to adapt it to spherical geometry.

In section \ref{background} we describe the particle-mesh
discretisation for the density equation. The method is modified to
form the remapped particle-mesh method in section \ref{remapped}.  We
discuss issues of efficient implementation in section \ref{efficient}.
In section \ref{numerics} we give numerical results for advection
tests in planar geometry and on the sphere, as well as results from
rotating shallow-water simulations using the remapped particle-mesh
method in the staggered leapfrog scheme \cite{reich_bit06}. 
We give a summary of our results and discussion in section \ref{summary}.

\ahead{Continuity equation and particle advection}
\label{background}
In this section we describe the particle-mesh discretisation for 
the density equation. This discretisation forms the basis for the
remapped particle-mesh method discussed in this paper. For simplicity,
we restrict the discussion to two-dimensional flows.

We begin with the continuity equation 
\begin{equation} \label{continuity}
\rho_t + \nabla\cdot(\rho {\bf u}) = 0,
\end{equation}
where $\rho$ is the density and ${\bf u} = (u,v)^T\in \mathbb{R}^2$ 
is the fluid velocity. We write (\ref{continuity}) 
in the Lagrangian formulation as
\begin{eqnarray}
\frac{D {\bf X}}{Dt} & = & {\bf u}, \label{hpm_mass1}\\
\rho({\bf x},t) &=& \int \rho_0({\bf a})\, \delta({\bf x} - 
{\bf X}({\bf a},t))\,{\rm d}A ({\bf a}), \label{hpm_mass2}
\end{eqnarray}
where $\rho({\bf x},t)$ is the density at time $t\ge 0$ at a 
fixed Eulerian position ${\bf x} = (x,y)^T \in \mathbb{R}^2$, 
\begin{equation}
\frac{D}{Dt}(\cdot) = (\cdot)_t + (\cdot)_x \,u + (\cdot)_y \,v
\end{equation}
is the Lagrangian time derivative, 
\begin{equation}
{\bf X}({\bf a},t)= (X({\bf a},t),Y({\bf a},t))^T \in \mathbb{R}^2
\end{equation}
is a Lagrangian particle position at time $t$ with initial position
${\bf X}({\bf a},0) = {\bf a} \in \mathbb{R}^2$, and $\rho_0({\bf a})
= \rho({\bf a},0)$ is the initial density at ${\bf a} = (a,b)^T \in
\mathbb{R}^2$.

To discretise the integral representation (\ref{hpm_mass2}), 
we introduce a finite set of Lagrangian particles ${\bf X}_\beta (t) 
= (X_\beta(t),Y_\beta(t))^T \in \mathbb{R}^2$, $\beta=1,\ldots,N$, and a
fixed Eulerian grid $\MM{x}_{k,l} = (x_k,y_l) = 
(k\cdot \Delta x,l \cdot \Delta y)^T$, 
$k,l=0,\ldots,M$. Then we approximate the Eulerian grid density 
$\rho_{k,l}(t) \approx \rho({\bf x}_{k,l},t)$ by 
\begin{equation} \label{hpm_mass3}
\rho_{k,l}(t) := \sum_\beta \rho_0({\bf a}_\beta)\,
\psi_{k,l}({\bf X}_\beta(t))\,{\rm d}A({\bf a}_\beta),
\end{equation}
where $\psi_{k,l}({\bf x})\ge 0$ are basis functions, which satisfy
$\int \psi_{k,l}({\bf x})\,{\rm d}A({\bf x}) = 1$. 
The initial particle positions
${\bf X}_\beta (0) = {\bf a}_\beta$ are assumed to form a grid and
${\rm d}A({\bf a}_\beta)$ is equal to the area of the associated 
grid cell. Equation (\ref{hpm_mass3}) may be simplified to
\begin{equation} \label{hpm_mass4}
\rho_{k,l}(t) = \sum_\beta m_\beta \,\psi_{k,l}({\bf X}_\beta (t)),
\end{equation}
where 
\begin{equation} \label{hpm_mass5}
m_\beta := \rho_0({\bf a}_\beta)\,{\rm d}A({\bf a}_\beta)
\end{equation}
is the ``mass'' of particle $\beta$.

Let us now also request that the basis functions $\psi_{kl}$ 
satisfy the partition-of-unity (PoU) property
\begin{equation}\label{PoU}
\sum_{k,l}\psi_{k,l}({\bf x}) \,{\rm d}A({\bf x}_{k,l}) = 1,\qquad
{\rm d}A({\bf x}_{k,l}) := \Delta x \Delta y,
\end{equation}
for all ${\bf x} \in \mathbb{R}^2$.
This ensures that the total mass is conserved since
\begin{equation}
\sum_{k,l} \rho_{k,l}(t)\,{\rm d}A({\bf x}_{k,l}) = 
\sum_{k,l} \sum_\beta m_\beta\,\psi_{k,l}({\bf X}_\beta(t))
\,{\rm d}A({\bf x}_{k,l}) = \sum_\beta
m_\beta,
\end{equation}
which is constant. The time evolution of the particle positions 
${\bf X}_\beta (t)$ is simply given by
\begin{equation} \label{hpm_mass6}
\frac{d}{dt} {\bf X}_\beta = {\bf u}_\beta.
\end{equation}

Given a time-dependent (Eulerian) velocity field ${\bf u}({\bf x},t)$,
we can discretise (\ref{hpm_mass5}) and (\ref{hpm_mass6}) 
in time with a simple differencing method:
\begin{eqnarray} \label{timestep1}
{\bf X}^{n+1}_\beta & = & {\bf X}^n_\beta+\Delta t {\bf u}_\beta^{n+1/2}, 
\qquad {\bf u}_\beta^{n+1/2} := {\bf u}({\bf X}_\beta^{n},t_{n+1/2}),\\
\rho_{k,l}^{n+1} & = & \sum_\beta m_\beta \,\psi_{k,l}({\bf X}_\beta^{n+1}).
\label{timestep2}
\end{eqnarray}
In \cite{frank02sp}, this discretisation was combined with a
time stepping method for the momentum equation to form a Hamiltonian
particle-mesh method for the rotating shallow-water equations. The masses
$m_\beta$ were kept constant throughout the simulation. In this paper, 
we instead combine the discretisation with a remapping technique 
so that the particles trajectories start from grid points 
at the beginning of each time step. Our remapping approach 
requires the assignment of new particle ``masses'' in each time step
and, hence, is fundamentally different from semi-Lagrangian 
remapping strategies described, for example, in \cite{Nair1}.

\ahead{Remapped particle-mesh method}
\label{remapped}

In this section, we describe the remapped particle-mesh method for
solving the continuity equation. The aim is to exploit the mass
conservation property of the particle-mesh method whilst keeping an
Eulerian grid data structure for velocity updates. To achieve this 
we reset the particles to an Eulerian grid point at the beginning
of each time step, i.e., 
\begin{equation}
{\bf X}_\beta^n := 
{\bf a}_\beta = {\bf x}_{k,l},\qquad \beta
= 1 + k + l\cdot M .
\end{equation}
This step requires the calculation of new particle ``masses'' 
$m_\beta^n$, $\beta = 1,\dots,N$, according to
\begin{equation}
\label{mat eqn}
\rho_{k,l}^n = \sum_\beta m_\beta^{n} \,\psi_{k,l}({\bf a}_\beta)
\end{equation}
for given densities $\rho_{k,l}^n$.
This is the remapping step. We finally step the particles 
forward and calculate the new density on the Eulerian grid 
using equations (\ref{timestep1})-(\ref{timestep2}) with
$m_\beta = m_\beta^n$. Note that the Lagrangian trajectory
calculation (\ref{timestep1}) can be replaced by any other
consistent upstream approximation. Exact trajectories for a given
time-independent velocity field ${\bf u}({\bf x})$ will, for example,
be used in the numerical experiments.

The whole process is mass conserving since the PoU property 
(\ref{PoU}) ensures that
\begin{equation}
\sum_{k,l}\rho_{k,l}^{n+1} \,{\rm d}A({\bf x}_{k,l}) = \sum_{k,l}
\sum_\beta m_\beta^n\,\psi_{k,l} ({\bf X}_\beta^{n+1})\,{\rm d}
A({\bf x}_{k,l})
= \sum_\beta m_\beta^n = \sum_{k,l} \rho_{k,l}^n \,{\rm d}A({\bf x}_{k,l}).
\end{equation}

\ahead{Efficient implementation}
\label{efficient}

This density advection scheme can be made efficient since all the
interpolation takes place on the grid; this means that the same
linear system of equations, characterized by (\ref{mat eqn}), 
is solved at each time step. The particle trajectories are uncoupled 
and thus may even be calculated in parallel.

The computation of the particle masses in (\ref{mat eqn}) 
leads to the solution of a sparse matrix system. We discuss
this issue in detail for (area-weighted) 
tensor product cubic $B$-spline basis functions,
defined by
\begin{equation} \label{cubicspline}
\psi_{k,l}({\bf x}) := \frac{1}{\Delta x \Delta y}\, 
\psi_{\rm cs}\left(\frac{x-x_{k}}{\Delta x}\right)
\cdot \psi_{\rm cs}\left(\frac{y-y_{l}}{\Delta y}\right),
\end{equation}
where $\psi_{\rm cs}(r)$ is the cubic B-spline
\begin{equation}
\psi_{\rm cs}(r) = \left\{ \begin{array}{ll}
\frac{2}{3} - |r|^2 + \frac{1}{2}|r|^3, & |r| \le 1,\\
\frac{1}{6}(2-|r|)^3, & 1 < |r| \le 2,\\
0, & |r| > 2. \end{array} \right.
\end{equation}
The basis functions satisfy
\begin{equation}
\sum_{k,l} \psi_{k,l} ({\bf x}) \,{\rm d}A({\bf x}_{k,l}) = 1
\end{equation}
and
\begin{equation}
\int \psi_{k,l}({\bf x})\,{\rm d}A({\bf x}) = 1
\end{equation}
as required.

A few basic manipulations reveal that (\ref{mat eqn}) becomes 
equivalent to
\begin{equation} \label{system}
\rho_{k,l}^n\,{\rm d}A({\bf x}_{kl}) =
\rho_{k,l}^n\,\Delta x \Delta y 
= \left(1 + \frac{\Delta x^2}{6} \delta_x^2 \right)
\, \left(1 + \frac{\Delta y^2}{6} \delta_y^2 \right)\, m_{k,l}^n
\end{equation}
where
\begin{equation}
\delta_x^2 \,m_{k,l}^n = \frac{m_{k+1,l}^n - 2m^n_{k,l} + m^n_{k-1,l}}{
\Delta x^2},\quad
\delta_y^2 \,m_{k,l}^n = \frac{m_{k,l+1}^n - 2m^n_{k,l} + m^n_{k,l-1}}{
\Delta y^2},
\end{equation}
are the standard second-order central difference approximations, and
we replaced index $\beta = 1+ k + l\cdot M$ by $k,l$, i.e., we write
$m^n_{k,l}$, ${\bf X}_{k,l}^n$, etc.~from now on. Eq.~(\ref{system})
implies that the particle masses can be found by solving a tridiagonal
system along each grid line (in each direction). 

If the cubic spline $\psi_{\rm cs}$ in (\ref{cubicspline}) 
is replaced by the linear spline
\begin{equation}
\psi_{\rm ls}(r) = \left\{ \begin{array}{ll}
1-|r|, & |r| \le 1,\\
0, & |r| > 1, \end{array} \right. 
\end{equation}
then the system (\ref{mat eqn}) is solved by 
\begin{equation}
m_{k,l}^n = \Delta x \Delta y\,\rho_{k,l}^n.
\end{equation}
The resulting low-order advection scheme possesses the desirable
property that $\rho_{k,l}^n \ge 0$ for all
$k,l$ implies that $\rho_{k,l}^{n+1} \ge 0$ for all $k,l$, and
so that monotonicity is also preserved. 

On a more abstract level, conservative advection schemes can be derived
for general (e.g.~triangular) meshes with basis functions 
$\phi_{kl}({\bf x})\ge 0$, which form a partition of unity. 
An appropriate quadrature formula for (\ref{hpm_mass2}) leads then 
to a discrete approximation of type (\ref{hpm_mass4}). This extension
will be the subject of a forthcoming publication.


\ahead{Extension to the sphere}
\label{sphere}
In this section we suggest a possible implementation of the remapped
particle-mesh method for the density equation on the sphere. The
method follows the particle-mesh discretisation given by
{\sc Frank \& Reich} \cite{FrRe03}, combined with 
a remapping to the grid.  

We introduce a longitude-latitude grid with equal grid spacing 
$\Delta \lambda = \Delta \theta = \pi/J$. 
The latitude grid points are offset
a half-grid length from the poles. Hence
we obtain grid points $(\lambda_k,\theta_l)$, where 
$\lambda_k = k\Delta \lambda$, 
$\theta_l = -\frac{\pi}{2} + (l-1/2)\Delta\theta$, 
$k=1,\dots,2J$, $l=1,\dots,J$,
and the grid dimension is $2J\times J$.

Let $\psi_{k,l}({\bf x})$ denote the (area-weighted) 
tensor product cubic
B-spline centered at a grid point ${\bf x}_{kl} \in \mathbb{R}^3$ with
longitude-latitude coordinates $(\lambda_k,\theta_l)$, 
i.e.
\begin{equation}\label{SPHbasis}
\psi_{k,l}({\bf x}) := \frac{1}{{\rm d}A({\bf x}_{k,l})}\,
\psi_{\rm cs} \left(\frac{\lambda-\lambda_k}{\Delta \lambda} \right)
\cdot  
\psi_{\rm cs} \left(\frac{\theta-\theta_l}{\Delta \theta} \right), 
\end{equation}
where $(\lambda,\theta)$ are the spherical coordinates of a point
${\bf x} = (x,y,z)^T \in \mathbb{R}^3$ on the sphere, $\psi_{\rm cs}(r)$
is the cubic B-spline as before, and
\begin{equation}
{\rm d}A({\bf x}_{k,l}) = R^2 
\cos(\theta_l)\,\Delta \theta \Delta \lambda .
\end{equation}
We convert between 
Cartesian and spherical coordinates using the formulas
\begin{equation}
x = R \cos \lambda \,\cos \theta,\quad y = R \sin \lambda \,\cos
\theta,\quad z = R \sin \theta,
\end{equation}
and
\begin{equation}
\lambda = \tan^{-1} \left( \frac{y}{x} \right),\qquad \theta =
\sin^{-1} \left( \frac{z}{R} \right).
\end{equation}

At each time step we write the fluid velocity in 3D Cartesian 
coordinates and step the particles ${\bf X}_{i,j}$ forward. 
We then project the particle positions onto the surface of the sphere
as described in \cite{FrRe03}. The Lagrangian trajectory algorithm is then:
\begin{equation}
{\bf X}^{n+1}_{i,j} = 
{\bf x}_{i,j} + \Delta t {\bf u}^{n+1/2}_{i,j} + \mu \,{\bf x}_{i,j},
\end{equation}
where $\mu$ is a Lagrange multiplier chosen so that
$\|{\bf X}^{n+1}_{i,j}\|=R$ on a sphere of radius $R$. This alogrithm
can be replaced by any other consistent approximation upstream
Lagrangian trajectories. Exact trajectories are, for example, used 
in the numerical experiments. 

We compute the particle masses $m_{i,j}^n$ by solving the system
\begin{equation} \label{sphere_advection}
\rho_{k,l}^n  = \sum_{i,j} 
m_{i,j}^n \,\psi_{k,l}({\bf x}_{i,j}) 
\end{equation}
for given densities $\rho_{k,l}^n$.
The density at time-level $t_{n+1}$ is then determined by
\begin{equation} \label{remapped_sphere}
\rho_{k,l}^{n+1}  = \sum_{i,j} 
m_{i,j}^n \,\psi_{k,l}({\bf X}_{i,j}^{n+1}).
\end{equation}
Note that the system (\ref{sphere_advection}) is equivalent to
\begin{equation} \label{sphere_advection2}
\rho_{k,l}^n \,{\rm d}A({\bf x}_{k,l}) 
= \left(1 + \frac{\Delta \lambda^2}{6} \delta_\lambda^2 \right)
\, \left(1 + \frac{\Delta \theta^2}{6} \delta_\theta^2 \right)\, m_{k,l}^n
\end{equation}
and can be solved efficiently as outlined in section \ref{efficient}.
The implementation of the remapping method is greatly simplified by
making use of the periodicity of the spherical coordinate
system in the following sense. The periodicity is trivial in the
longitudinal direction. For the latitude, a great circle meridian
is formed by connecting the latitude data separated by an angular
distance $\pi$ in longitude (or $J$ grid points). See, for
example, the paper by {\sc Spotz, Taylor \& Swarztrauber}
\cite{spotz}. It is then efficient to solve the system 
(\ref{sphere_advection2}) using a direct solver.

Conservation of mass is encoded in
\begin{equation}
\sum_{k,l} \rho_{k,l}^{n+1}\, {\rm d}A({\bf x}_{k,l}) = \sum_{k,l} 
\rho_{k,l}^{n} \,{\rm d}A({\bf x}_{k,l}),
\end{equation}
which holds because of the PoU property
\begin{equation}
\sum_{k,l} \psi_{kl}({\bf x}) \,{\rm d}A({\bf x}_{k,l}) = 1.
\end{equation}


\ahead{Numerical results} 
\label{numerics}

\bhead{1D convergence test} 
Following \cite{Zerroukat3}, we test the convergence rate of our method for 
one-dimensional uniform advection of a sine wave over a periodic 
domain $\Omega = [0,1)$. The initial distribution is
\begin{equation}
\rho_0(x) = \sin (2\pi x)
\end{equation}
and the velocity field is $u(x,t) = U = 1$. 
The 1D version of our method is used to solve the continuity equation
\begin{equation}
\rho_t = -(\rho u)_x.
\end{equation}
The experimental setting is equivalent to that of \cite{Zerroukat3}. Table
\ref{table1} displays the convergence of $l_2$ errors as a function of 
resolution $\Delta x = 1/M$. Note that the results from Table \ref{table1}
are in exact agreement with those displayed in Table I of \cite{Zerroukat3}
for the parabolic spline method (PSM) and fourth-order accuracy
is observed.

\begin{table} \begin{center}
\begin{tabular}{|c|ccccccc|}
\hline & & & & & & & \\ 
$M$ & 8 & 16 & 32 & 64 & 128 & 256 & 512 \\
\hline & & & & & & & \\
$l_2$ & 0.549E-02 & 0.254E-03 & 0.143E-4 & 0.872E-6 & 0.541E-07 &
0.337E-08 & 0.211E-09 \\
\hline 
\end{tabular} \end{center}
\caption{Convergence of $l_2$-errors as a function of $\Delta x = 1/M$
for uniform advection with $U=1$ of a sine wave on a periodic 
domain $\Omega = [0,1)$ 
with $\Delta t = 0.12 \Delta x/U$ and 20 time steps.}
\label{table1}
\end{table}

\begin{figure} 
\begin{center}$\,  ${Cubic Interpolation}$\qquad \qquad \qquad \qquad 
\quad  ${Cubic Spline} \qquad\end{center}

\begin{center}\includegraphics[  scale=0.3]{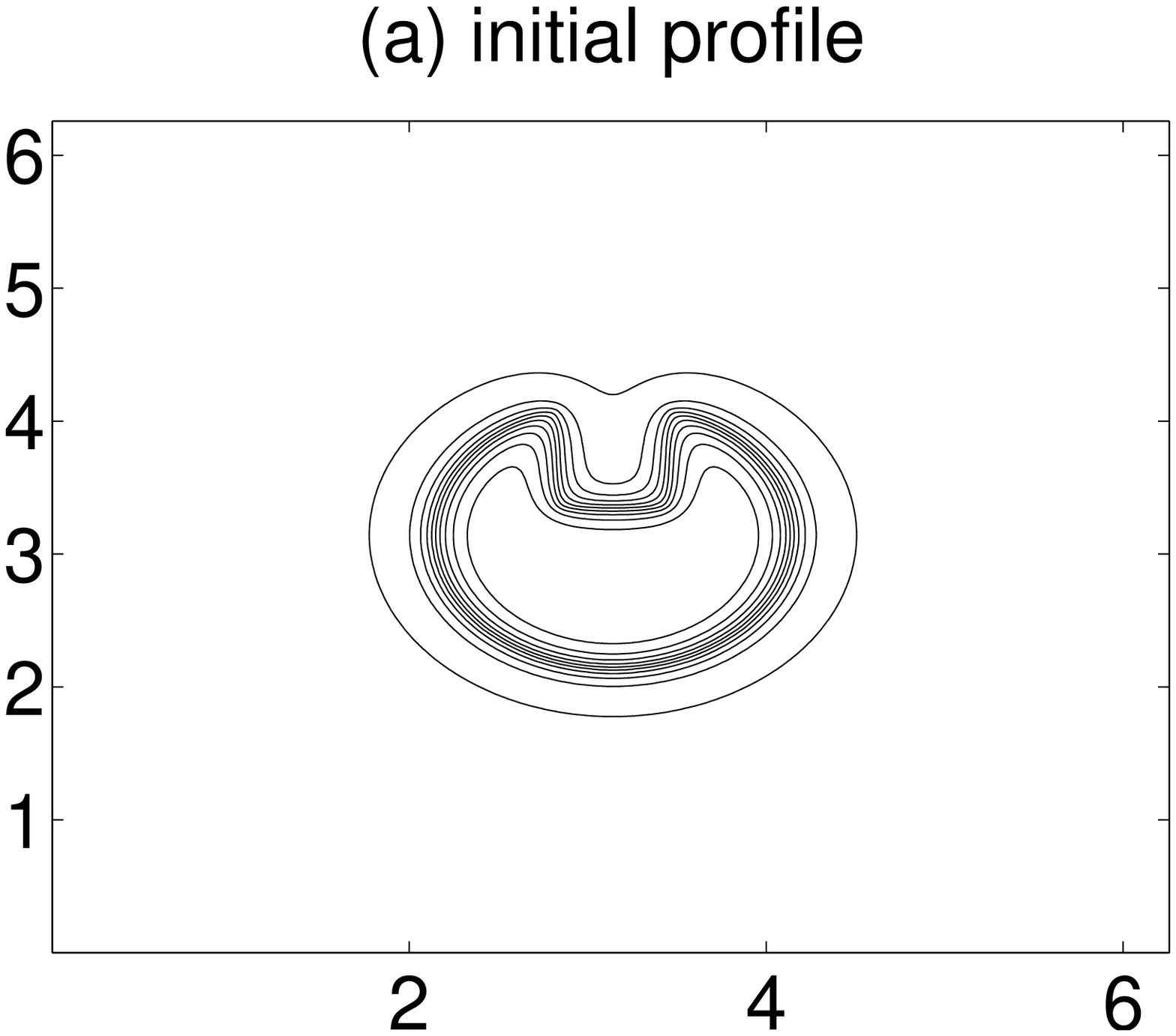}$\qquad$
\includegraphics[  scale=0.3]{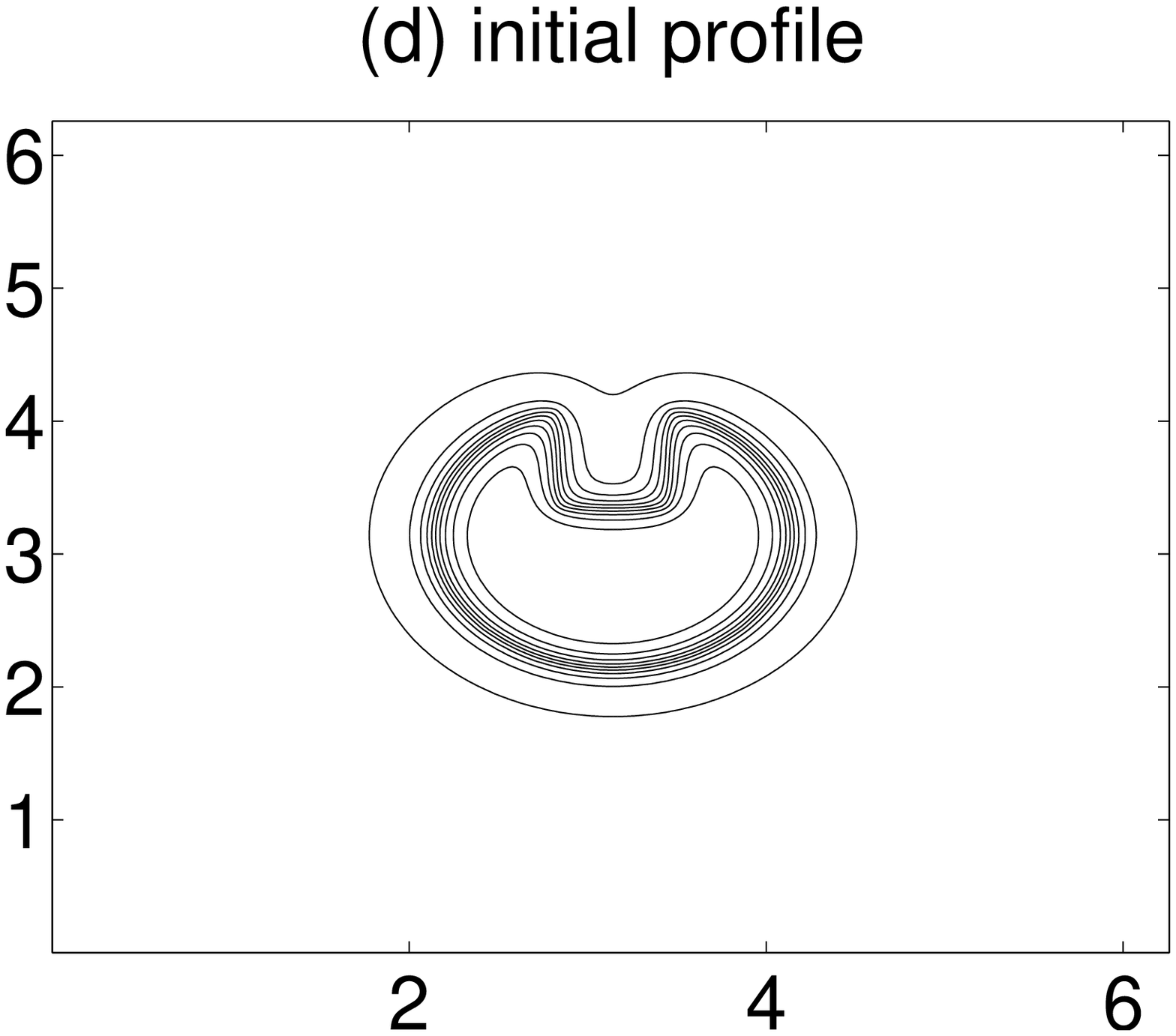}\end{center}

\begin{center}\includegraphics[  scale=0.3]{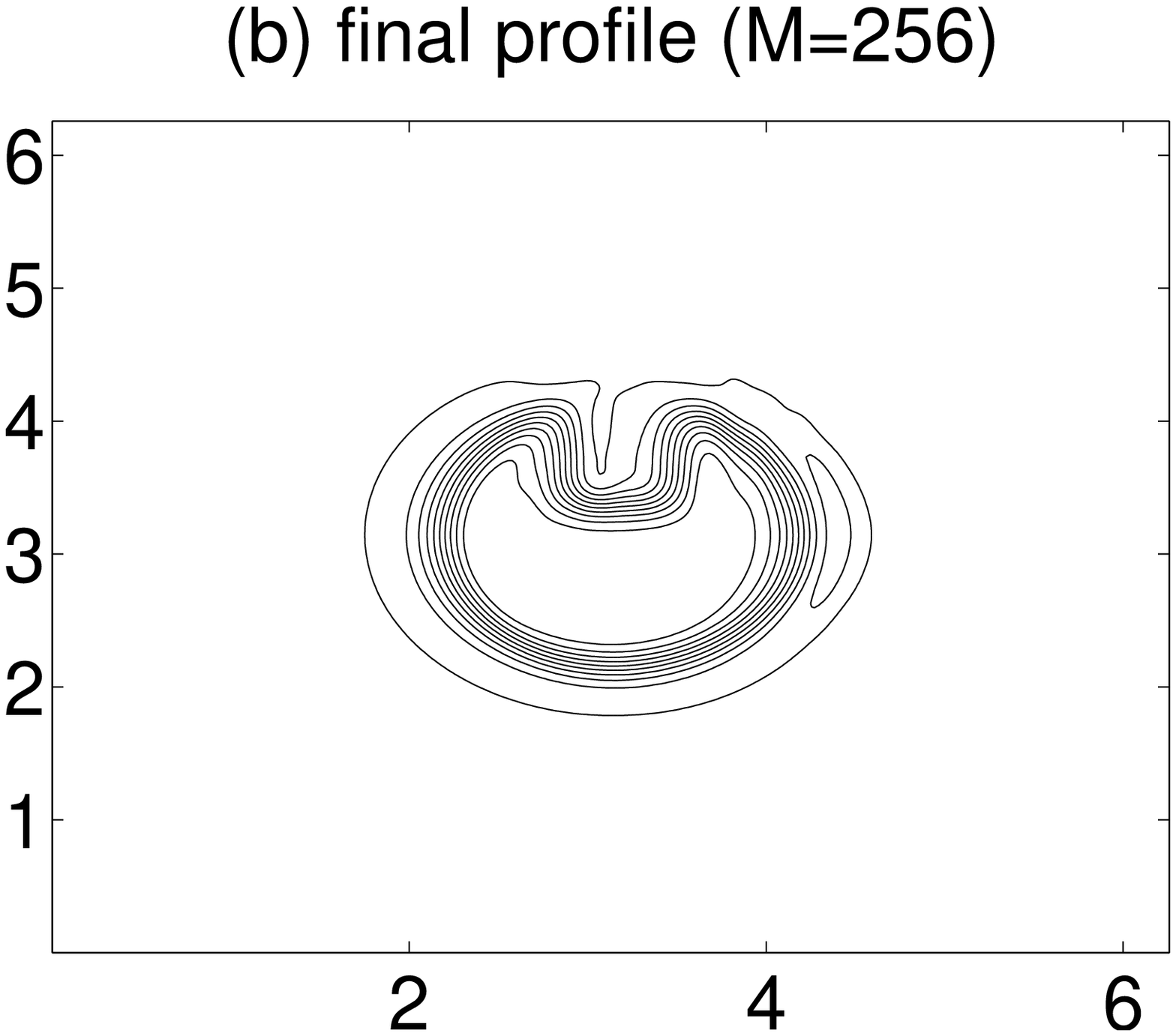}$\qquad$
\includegraphics[  scale=0.3]{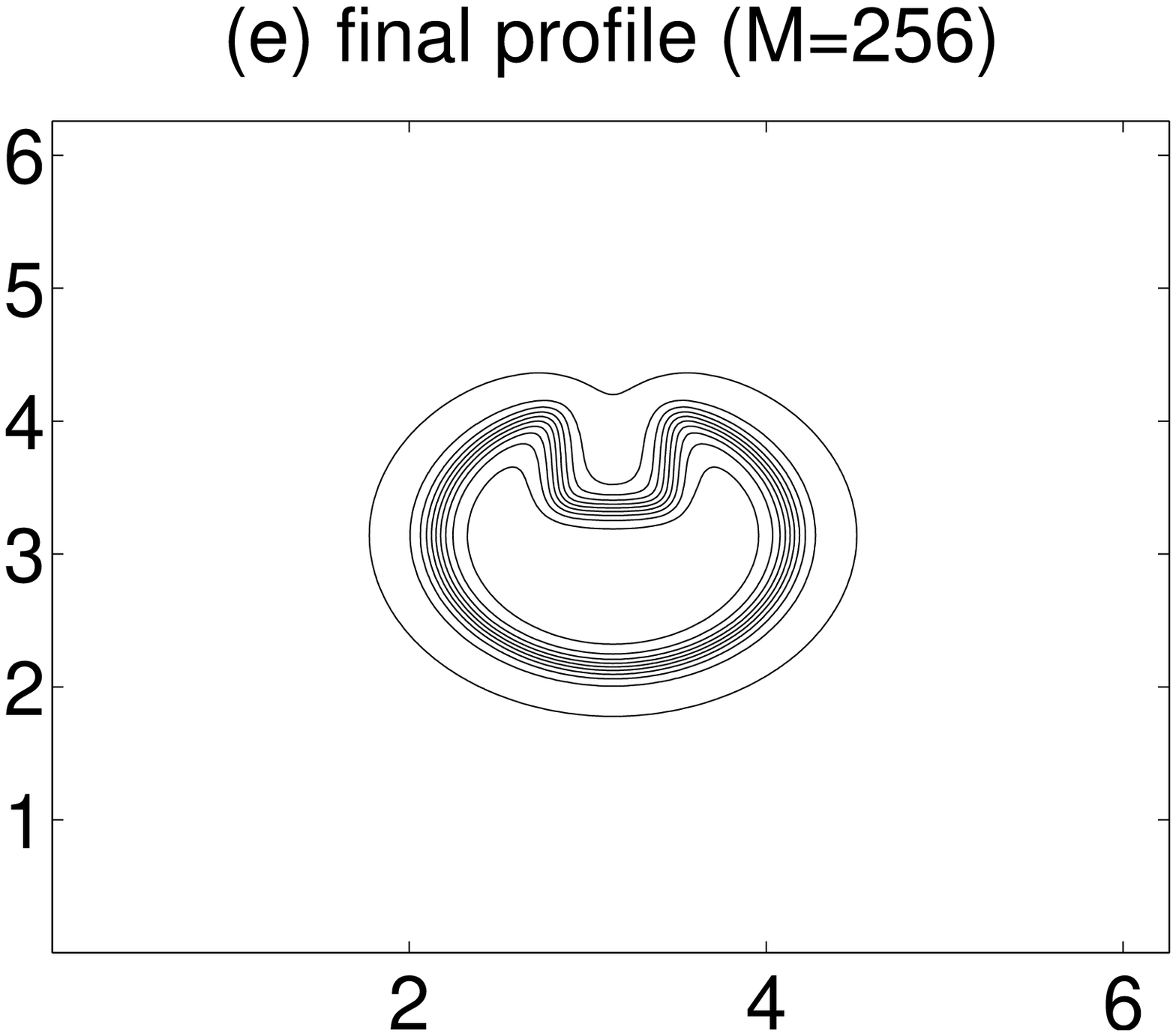}\end{center}

\begin{center}\includegraphics[  scale=0.3]{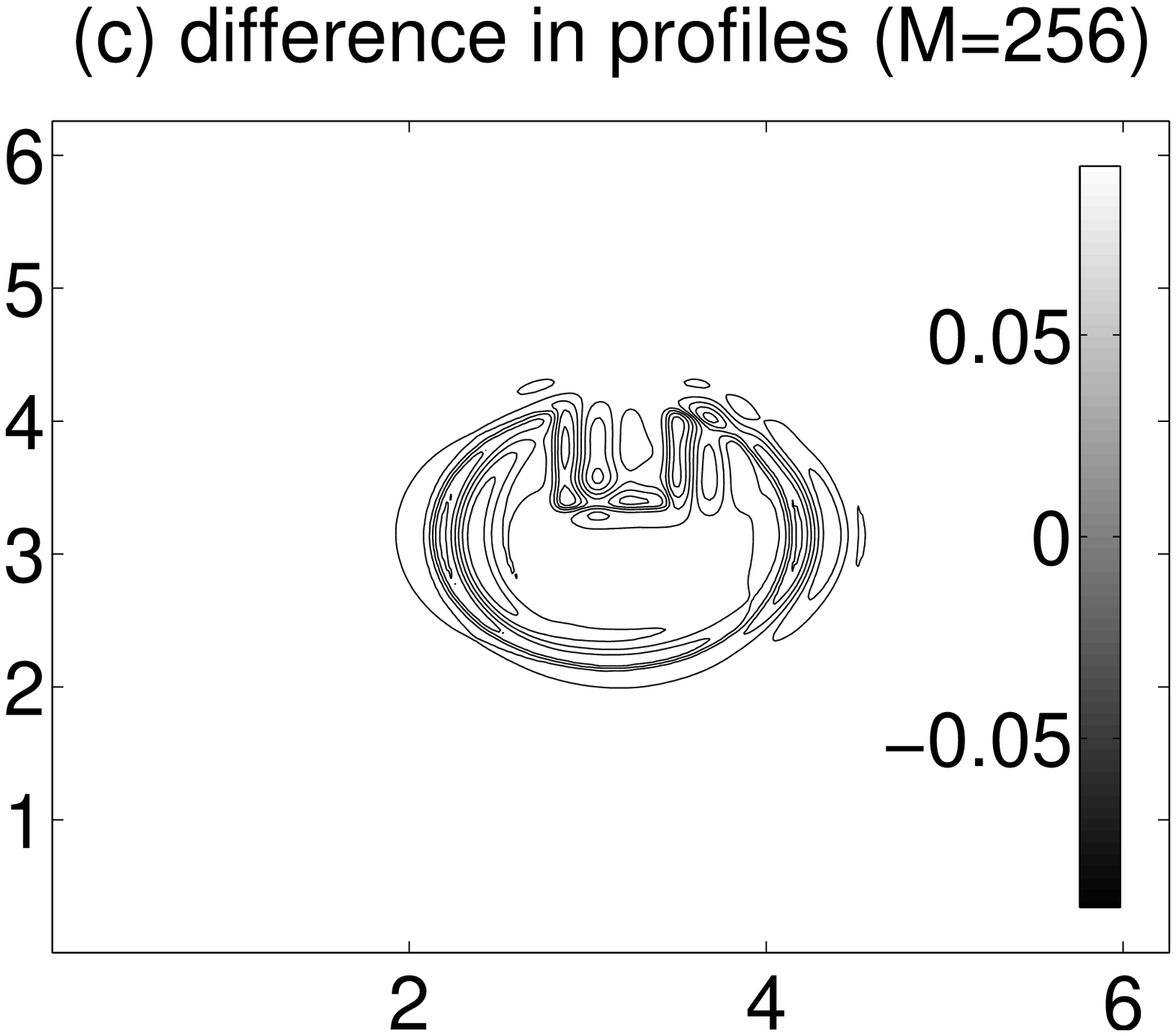}$\qquad$
\includegraphics[  scale=0.3]{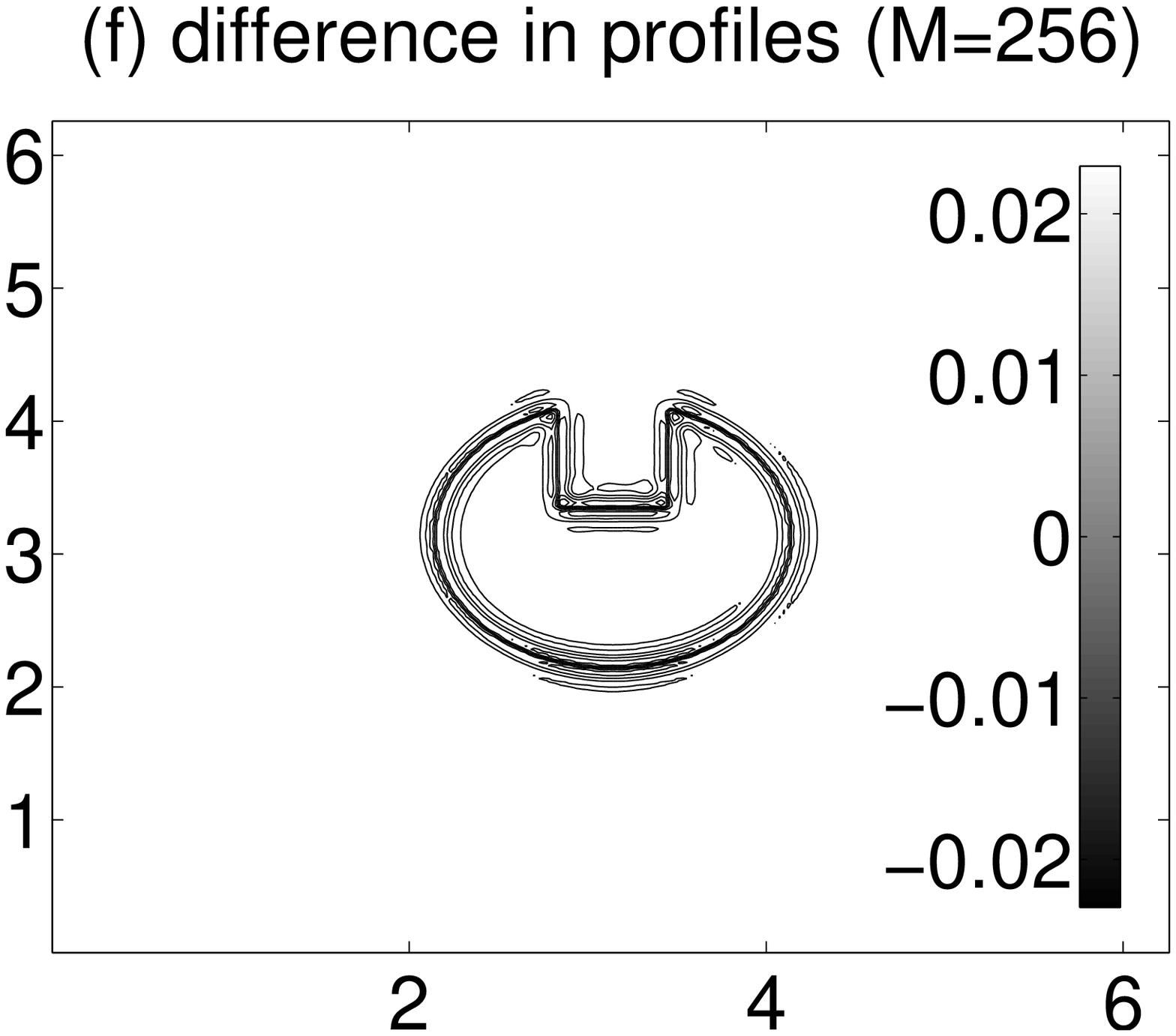}\end{center}

\caption{\label{cap:newfigure1a} Detailed results from the 
linearly advected smoothed
slotted-cylinder experiment with $M=256$. Left panels:
 classic SL interpolation using backward trajectories
and bicubic interpolation. Right panels: new advection scheme using forward
trajectories and mass-conserving spline interpolation.}
\label{figure1a}
\end{figure}

\begin{figure} 

\begin{center}\includegraphics[  scale=0.32]{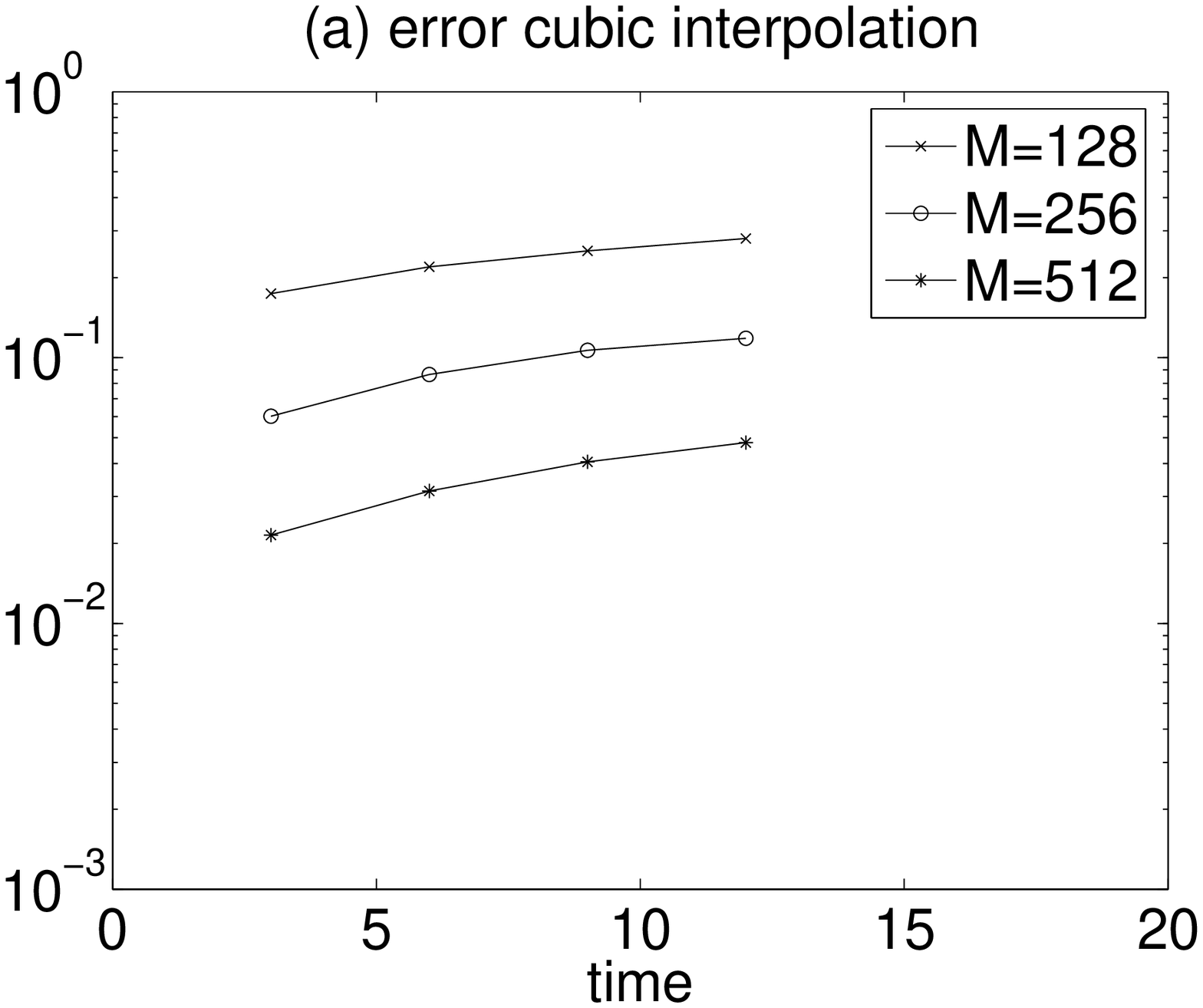}$\qquad$
\includegraphics[  scale=0.32]{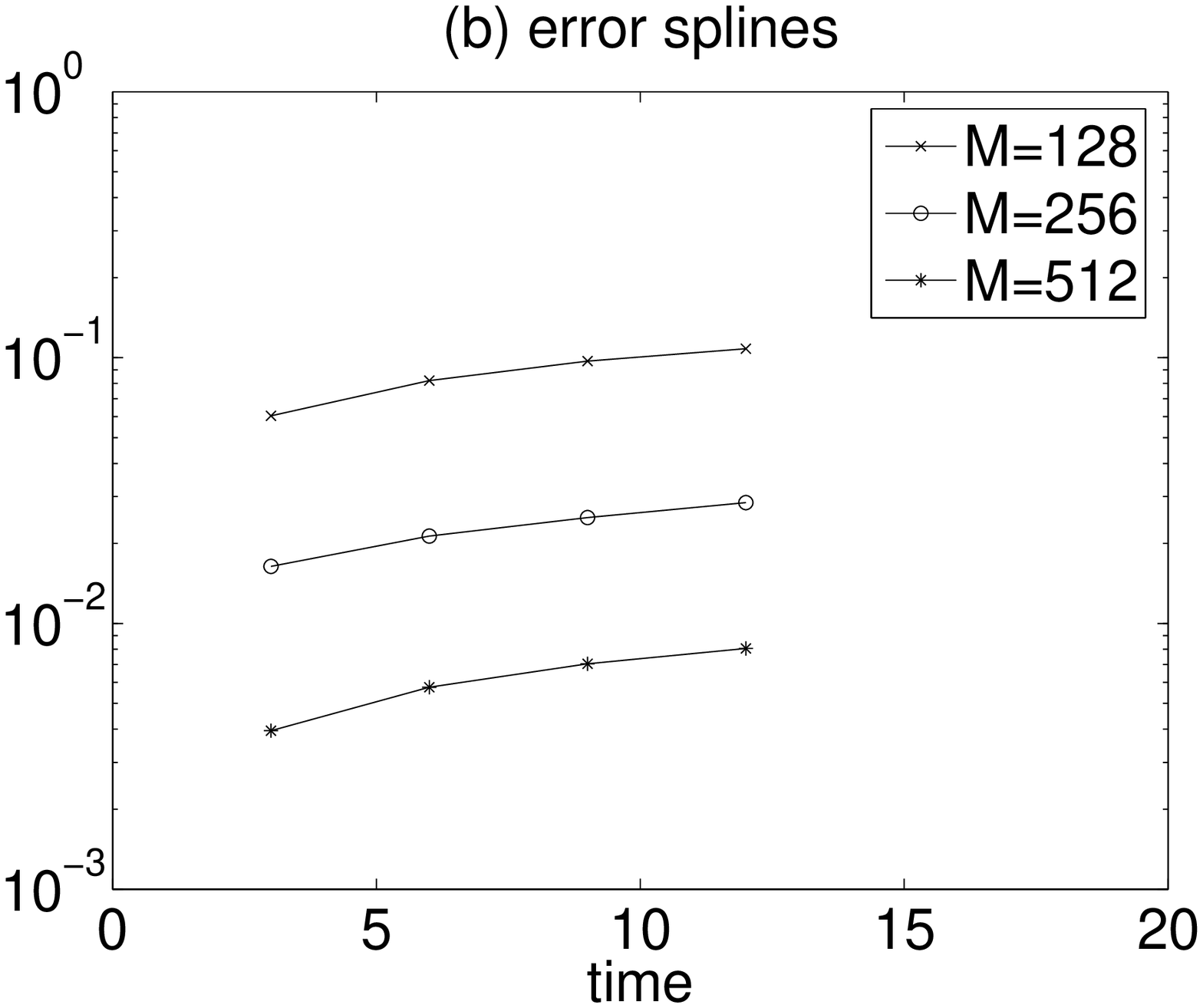}\end{center}

\caption{\label{cap:newfigure1b} Displayed are the 
$l_\infty$-errors (\ref{error}) for constant step-size
$\Delta t$ and varying spatial resolution for the 
lineearly advected smoothed slotted
cylinder experiment. 
Left panel: classic SL interpolation using backward trajectories
and bicubic interpolation. Right panel: new advection scheme using forward
trajectories and mass-conserving spline interpolation.}
\label{figure1b}
\end{figure}

\bhead{2D planar advection: Slotted-cylinder problem}

Convergence is now examined for a more realistic test case.
Since we use higher-order interpolation the initial
density profile $\rho_0$ needs to be sufficiently smooth. On the other hand, 
relatively sharp gradients should be present to pose a challenge to the
advection scheme. We decided to use a smoothed slotted-cylinder obtained
by applying a modified Helmholtz operator ${\cal H} = 
(I - \alpha^2 \nabla^2)^{-1}$
to the standard sharp-edged slotted cylinder \cite{Zalesak}. The smoothing 
length is set to $\alpha = 2\pi/64$. See panel (a) in 
Fig.~\ref{figure1a}.

We compare the newly proposed scheme to the standard SL advection
scheme based on backward trajectories and bicubic interpolation
(see, e.g.~\cite{Staniforth91}). To exclude any errors from the
trajectory calculation we use a double periodic domain of size
$[0 ,2\pi] \times [0,2\pi]$ and apply a constant velocity field
$u = 2\pi/3$, $v = 2\pi$. The time-step is $\Delta t = 0.01$ and
the simulations are run over a period of $T = 12$ time units. Note
that the initial density profile $\rho_0$ returns to its original position
after $\tau = 3$ time units. This allows us to introduce the error
\begin{equation} \label{error}
e_m = \|\rho_0({\bf x}_{k,l}) - 
\rho_{k,l}^{m K}\|_\infty, \qquad K = 300,
\end{equation}
for $m=1,2,3,4$. 

Simulations are performed on a spatial grid with $M = 128$, $M=256$,
and $M=512$. Errors (\ref{error}) are provided in
Fig.~\ref{figure1b}. It can be seen that the newly proposed method
is more accurate than the standard SL advection scheme and that
the newly proposed method achieves second-order accuracy as a function
of spatial resolution (for fixed time-steps $\Delta t$). Detailed
results from simulations with $M=256$ can be found in Fig.~\ref{figure1a}.

Following the discussion of \cite{Zerroukat3} the reduced order can be
explained by the fact that the Helmholtz operator ${\cal H}$ leads to
an approximate $|k|^{-2}$ spectral decay in the Fourier transform
of the initial density $\rho_0$. As also explained in \cite{Zerroukat3},
the improved convergence of our spline-based method over the traditional
bicubic SL method is to be expected.  

\begin{figure}
\begin{center}\includegraphics[  scale=0.5]{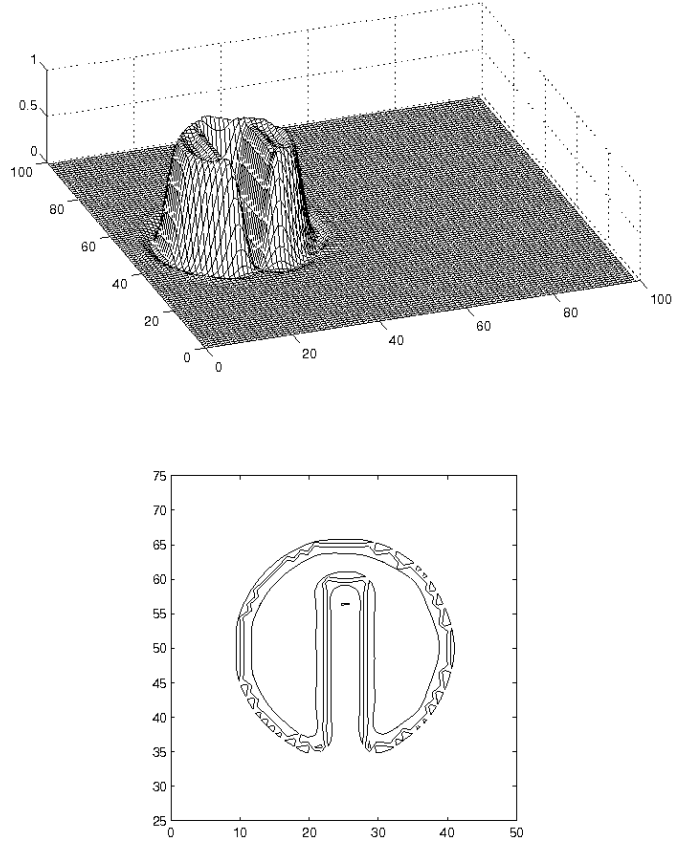}
\end{center}
\caption{Rotating slotted-cylinder problem. Top panel: numerical solutation 
after six rotations. 
Bottom panel: error (analytic minus numerical) with contour
minimum $-0.5266$ and contour interval $0.3803$; error measures,
as defined in \cite{Zerroukat1}, rms$_1$ = 0.062595,
rms$_2$ = 0.037329, and pdm = -0.1454E-10\,\%.}
\label{figure7}
\end{figure}

We also implemented the standard rotating 
slotted-cylinder problem as, for example,
defined in \cite{NCS99b,Zerroukat1}. See \cite{Zerroukat1} for a
detailed problem description and numerical reference solutions. 
Corresponding results for the newly proposed advection scheme 
can be found in Fig.~\ref{figure7}.


\begin{figure}
\begin{center}\includegraphics[  scale=0.5]{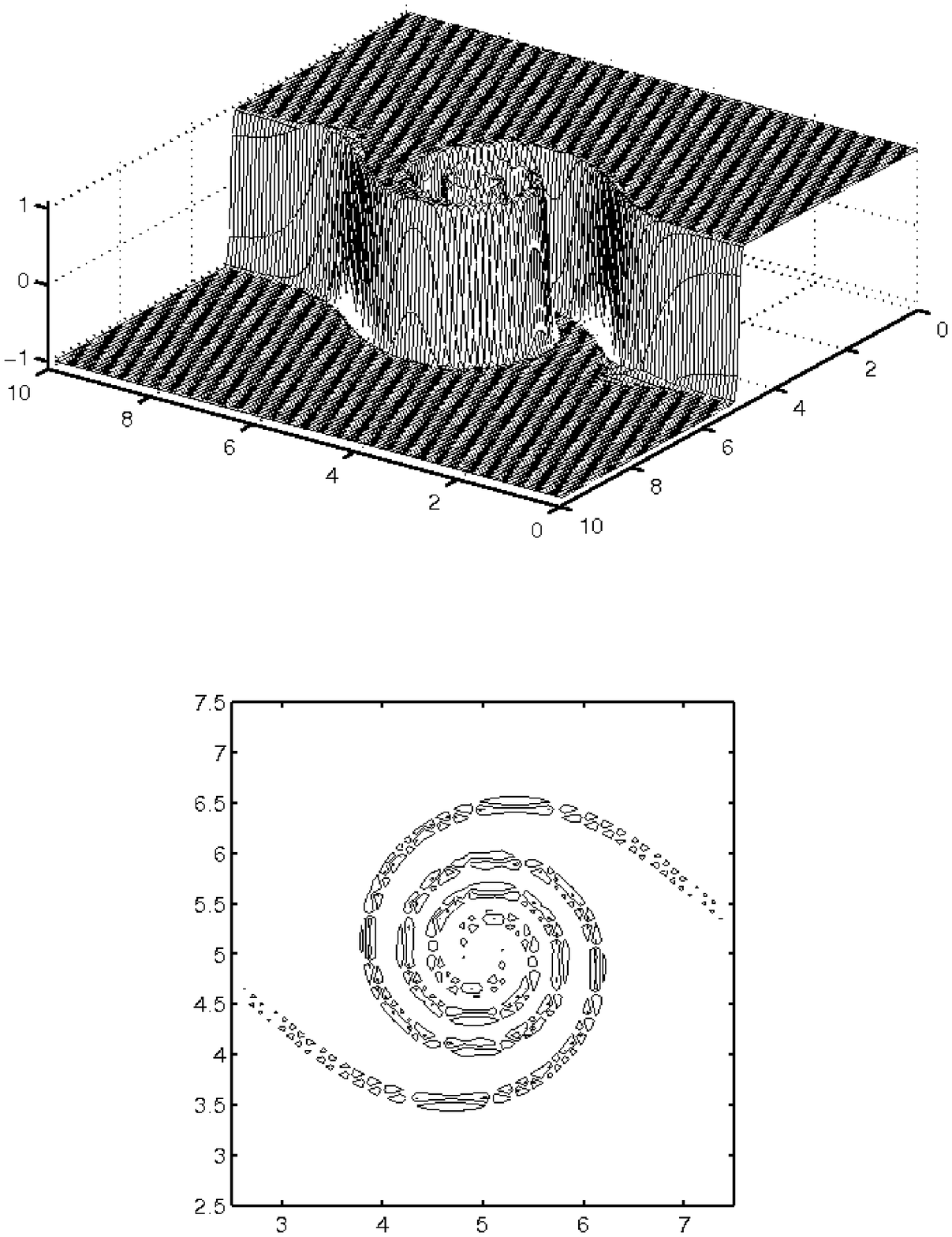}
\end{center}
\caption{Cyclogenesis problem. Top panel: numerical soluation at time
$t = 5$. Bottom panel: error (analytic minus numerical) with contour
minimum $-0.627$ and contour interval $0.418$; error measures,
as defined in \cite{Zerroukat1}, rms$_1$ = 0.081439,
rms$_2$ = 0.037703, and pdm = -0.176259E-11\,\%.}
\label{figure6}
\end{figure}

\bhead{2D planar advection: Idealized cyclogenesis problem}

The idealized cyclogenesis problem (see, e.g., \cite{NCS99b,Zerroukat1})
consists of a circular vortex with a tangential velocity $V(r) = 
v_0\,\tanh(r)/\mbox{sech}^2(r)$, where $r$ is the radial distance from the
centre of the vortex $(x_c,y_c)$ and $v_0$ is a constant chosen such
that the maximum value of $V(r)$ is unity. The analytic solution
$\rho({\bf x},t)$ is
\begin{equation}
\rho({\bf x},t) = -\tanh \left[ \left( \frac{y-y_c}{\delta} \right)
\cos(\omega t) - \left( \frac{x-x_c}{\delta} \right) \sin (\omega t)
\right],
\end{equation}
where $\omega = V(r)/r$ is the angular velocity and $\delta = 0.05$.
The experimental setting is that of \cite{NCS99b,Zerroukat1}. In particular,
the domain of integration is $\Omega = [0,10]\times [0,10]$ with
a $129\times 129$ grid. The time 
step is $\Delta t = 0.3125$ and a total of 16 time steps is performed. 
Numerical reference solutions can be found in \cite{Zerroukat1} for
the standard bicubic and several conservative SL methods. The corresponding 
results for the newly proposed advection scheme can be found 
in Fig.~\ref{figure6}.


\bhead{Spherical advection: Solid body rotation}

Solid body rotation is a commonly used experiment to test an advection
scheme over the sphere. We apply the experimental setting of
\cite{Nair1,Nair2,Nair3,Zerroukat2}. The initial density is the
cosine bell,
\begin{equation}
\rho_0(\lambda,\theta) = \left\{ \begin{array}{lr} 
1/2\,[1 + \cos(\pi r/R)], & r \le R,\\
0, & r > R, \end{array} \right.
\end{equation}
where $R = 7\pi/64$,
\begin{equation}
r = \cos^{-1} \left[ \sin \theta + \cos \theta\,
\cos(\lambda - \lambda_c) \right],
\end{equation}
and $\lambda_c = 3\pi/2$. The bell is advected by a time-invariant
velocity field
\begin{eqnarray}
u &=& \cos \alpha\,\cos \theta + \sin \alpha \, \cos \lambda \, \sin \theta,\\
v &=& -\sin \alpha\,\sin \lambda,
\end{eqnarray}
where $(u,v)$ are the velocity components in $\lambda$ and $\theta$ direction,
respectively, and $\alpha$ is the angle between the axis of solid body rotation
and the polar axis of the sphere. 

\begin{table} \begin{center}
\begin{tabular}{|c|c|c|c|}
\hline & &  &  \\ 
$\alpha$ & $0$ & $\pi/2$ & $\pi/2-0.05$  \\
\hline & &  &  \\
$l_1$ & 0.0492 & 0.0591 & 0.0627  \\
\hline & & &    \\
$l_2$ & 0.0336 & 0.0393 & 0.0397  \\
\hline & &  & \\
$l_\infty$ & 0.0280 & 0.0367 & 0.0374  \\
\hline 
\end{tabular} \end{center}
\caption{Comparison of error norms for solid body rotation with 
three different values of $\alpha$ $t$ after one complete revolution
using 256 time steps over a $128 \times 64$ grid. The meridional 
Courant number is $C_\theta = 0.5$.}
\label{table3}
\end{table}

Experiments are conducted for
$\alpha = 0$, $\alpha = \pi/2$, and $\alpha = \pi/2-0.05$. Analytic
trajectories are used and $\Delta t$ is chosen such that 256 time steps
correspond to a complete revolution around the globe (the radius of the
sphere is set equal to one). Accuracy is measured as relative errors
in the $l_1$, $l_2$, and $l_\infty$ norms (as defined, for example, in
\cite{Zerroukat2}). Results are reported in
Table \ref{table3} for a $128 \times 64$ grid (i.e., $J=64$). 

Note that (\ref{sphere_advection2}) may lead to a non-uniform
distribution of particle masses near the polar cap regions for
meridional Courant numbers $C_\theta > 1$. This can
imply a loss of accuracy if a ``heavy'' extra-polar particle
moves into a polar cap region. We verified this for 72, 36 and 18, 
respectively, time steps per complete revolution (implying a 
meridional Courant number of $C_\theta = 1.78$, $C_\theta = 3.56$,
and $C-\theta = 7.12$, respectively). 
It was found that the accuracy is improved by 
applying a smoothing operator along lines 
of constant $\theta$ near the polar caps, e.g.,
\begin{equation} \label{smoothing}
\rho^{n+1} = \left[ 1 - \left(\frac{\beta}{\cos \theta}\right)^6
\frac{\partial^6}{\partial \lambda^6} \right]^{-1} \rho^{n+1}_\ast ,
\end{equation}
$\beta \ll \pi/J$, $J = 64$. Here $\rho^{n+1}_\ast$ denotes
the density approximation obtained from (\ref{remapped_sphere}). The filter
(\ref{smoothing}) is mass conserving and acts similarly to
hyper-viscosity. The disadvantage of 
this simple filter is that $\rho^{n+1} \not= \rho^n$ 
under zero advection.

\begin{table} \begin{center}
(a) 72 time steps $  \qquad \quad \quad $ (b) 
36 time steps $ \qquad  \qquad  $ (c) 18 time steps
\vspace{0.5cm} 

\begin{tabular}{|c|c|c|}
\hline & &    \\ 
$\beta$ & $0$ & $\pi/(3J)$    \\
\hline & &    \\
$l_1$ & 0.0491 & 0.0283   \\
\hline & &     \\
$l_2$ & 0.0468 & 0.0168   \\
\hline & &   \\
$l_\infty$ & 0.0723 & 0.0122   \\
\hline 
\end{tabular} $  \qquad $ 
\begin{tabular}{|c|c|c|}
\hline & &    \\ 
$\beta$ & $0$ & $\pi/(3J)$    \\
\hline & &    \\
$l_1$ & 2.3264 & 0.0222   \\
\hline & &     \\
$l_2$ & 1.5124 & 0.0137   \\
\hline & &   \\
$l_\infty$ & 1.1383 & 0.0151   \\
\hline 
\end{tabular} 
$ \qquad $ 
\begin{tabular}{|c|c|c|}
\hline & &    \\ 
$\beta$ & $0$ & $\pi/(3J)$    \\
\hline & &    \\
$l_1$ & 2.3217 & 0.0143   \\
\hline & &     \\
$l_2$ & 1.5126 & 0.0105   \\
\hline & &   \\
$l_\infty$ & 1.0764 & 0.0143   \\
\hline 
\end{tabular} 
\end{center}
\caption{Comparison of error norms for solid body rotation with 
$\alpha=\pi/2$  for different values of the smoothing
parameter $\beta$ in (\ref{smoothing}) after one complete revolution
over a $128 \times 64$ grid (i.e., $J=64$). Panel (a): 
Complete revolution using 72 time step. The meridional 
Courant number is $C_\theta = 1.78$. Panel (b): 
Complete revolution using 36 time step. The meridional 
Courant number is $C_\theta = 3.56$. Panel (c): 
Complete revolution using 18 time step. The meridional 
Courant number is $C_\theta = 7.12$.

}
\label{table4}
\end{table}

Results for $\beta = 0$ and $\beta = \pi/192$,
respectively, and 72, 36 and 18 time steps, respectively, 
are reported in Table \ref{table4}. It is evident that
filtering by (\ref{smoothing}) improves the results
significantly. Corresponding results for
standard advection schemes can be found in \cite{Nair1} for
the case of 72 time steps per complete revolution. 


\bhead{Spherical advection: Smooth deformational flow}

To further evaluate the accuracy of the advection scheme in spherical
geometry, we consider the idealized vortex problem of
{\sc Doswell} \cite{Doswell}.
The flow field is deformational and an analytic solution is available
(see \cite{NCS99,Nair1} for details). 

We summarize the mathematical formulation. 
Let $(\lambda',\theta')$ be a rotated coordinate system with
the north pole at $(\pi+0.025,\pi/2.2)$ with respect to
the regular spherical coordinates. We consider rotations
of the  $(\lambda',\theta')$ coordinate system with an angular
velocity $\omega$, i.e.,
\begin{equation}
\frac{d\lambda'}{dt} = \omega, \qquad
\frac{d\theta'}{dt} = 0,
\end{equation}
where
\begin{equation}
\omega (\theta') = \frac{3\sqrt{3}\, \mbox{sech}^2(3\cos \theta') 
\tanh (3 \cos \theta')}{
6 \cos \theta'} .
\end{equation}
An analytic solution to the continuity equation 
(\ref{continuity}) in $(\lambda',\theta')$ 
coordinates is provided by
\begin{equation}
\rho(\lambda',\theta',t) = 1-\tanh \left[
\frac{3 \cos \theta'}{5} \sin(\lambda' - \omega(\theta')\, t)\right].
\end{equation}

\begin{table} \begin{center}
\begin{tabular}{|c|c|c|}
\hline & &  \\ 
$t$ & 3 & 6  \\
\hline & &  \\
$l_1$ & 0.0019 & 0.0055  \\
\hline & &  \\
$l_2$ & 0.0062 & 0.0172  \\
\hline & &  \\
$l_\infty$ & 0.0324 & 0.0792  \\
\hline 
\end{tabular} \end{center}
\caption{Comparison of error norms at different times $t$ for
spherical polar vortex problem. 
Computations are performed with a step size of 
$\Delta t = 1/20$ and a $128 \times 64$ grid.}
\label{table2}
\end{table}

\begin{figure}

\begin{center}\includegraphics[  scale=0.4]{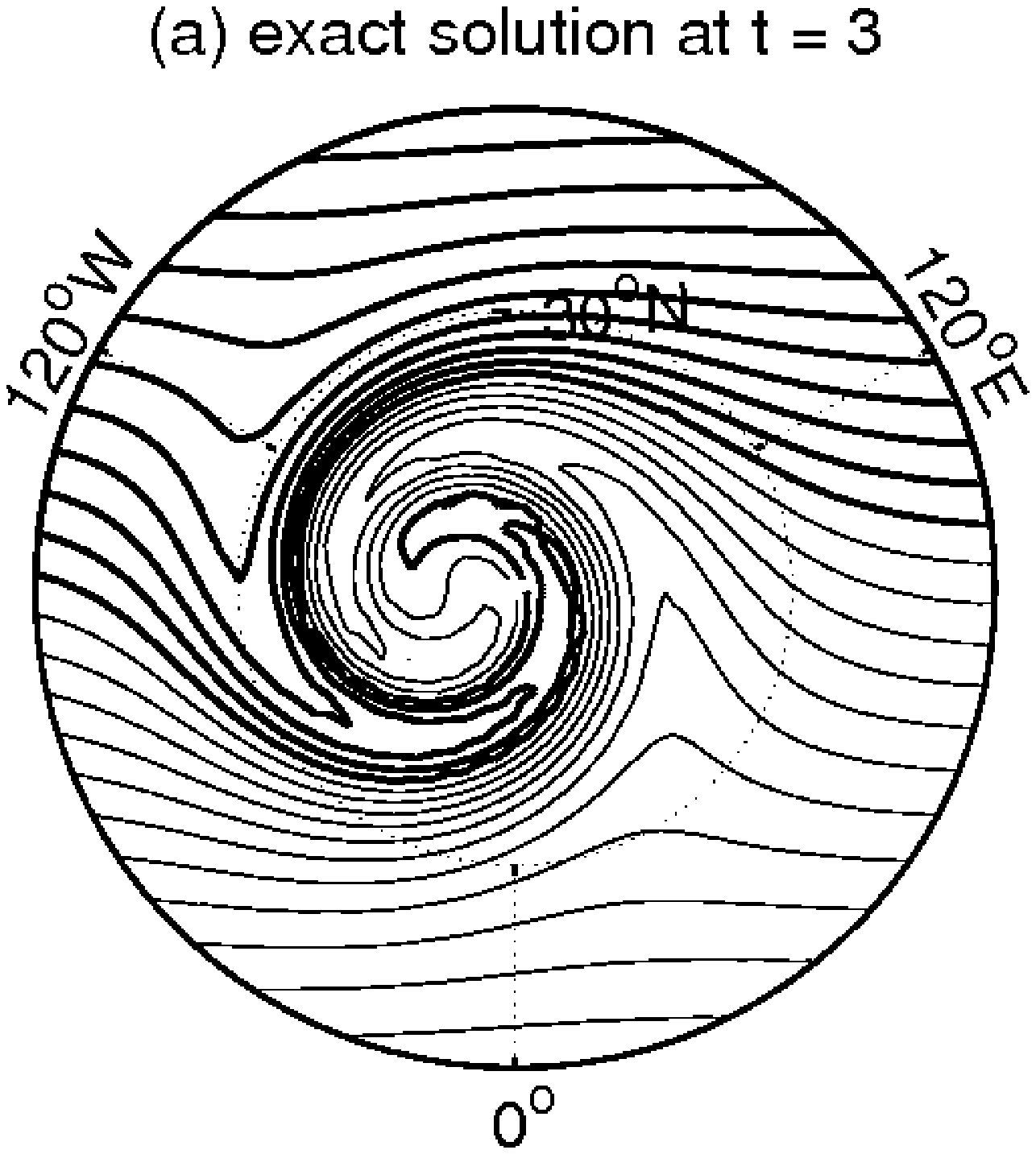}
$\qquad$\includegraphics[  scale=0.4]{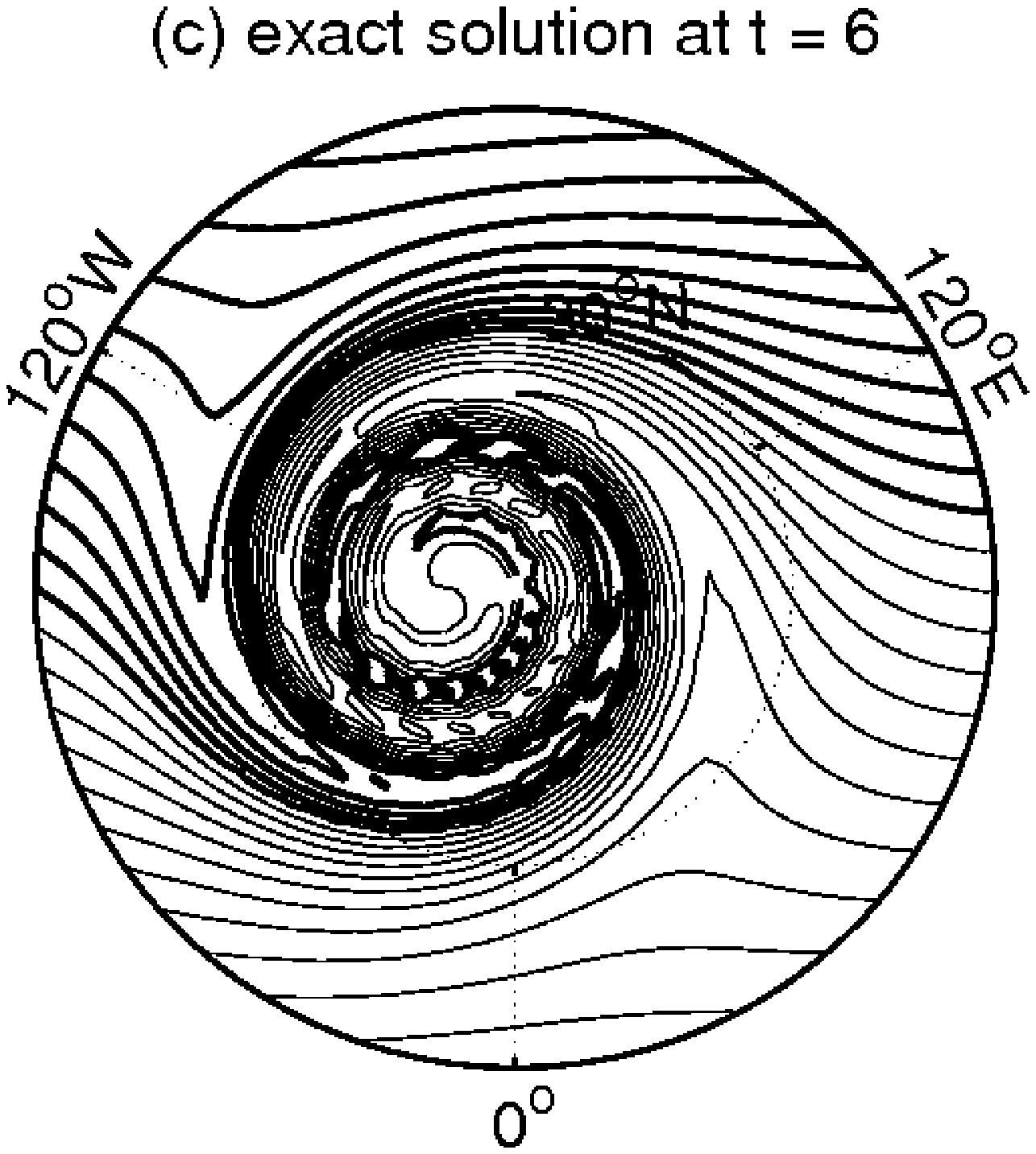}\end{center}

\begin{center}\includegraphics[  scale=0.4]{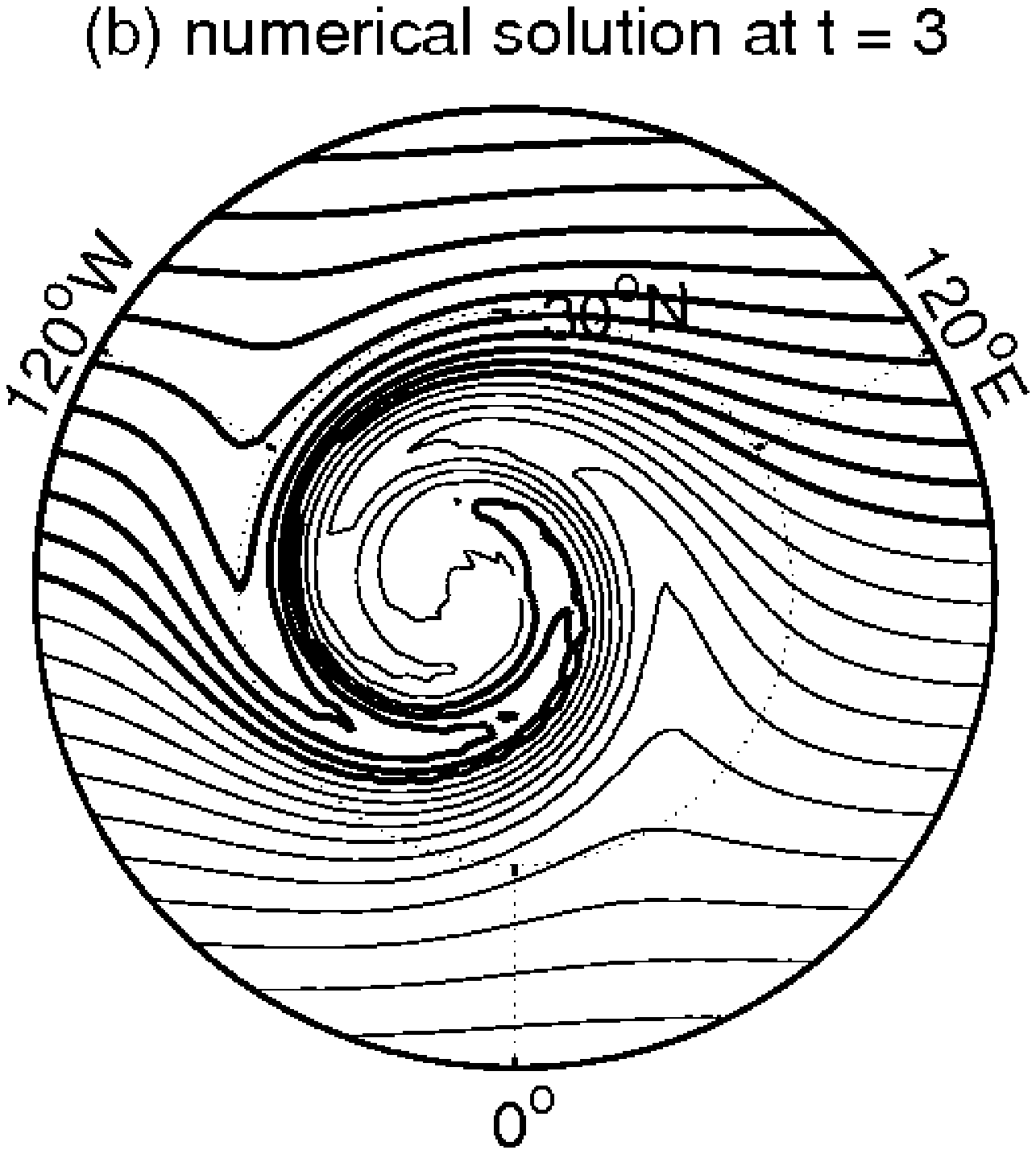}
$\qquad$\includegraphics[  scale=0.4]{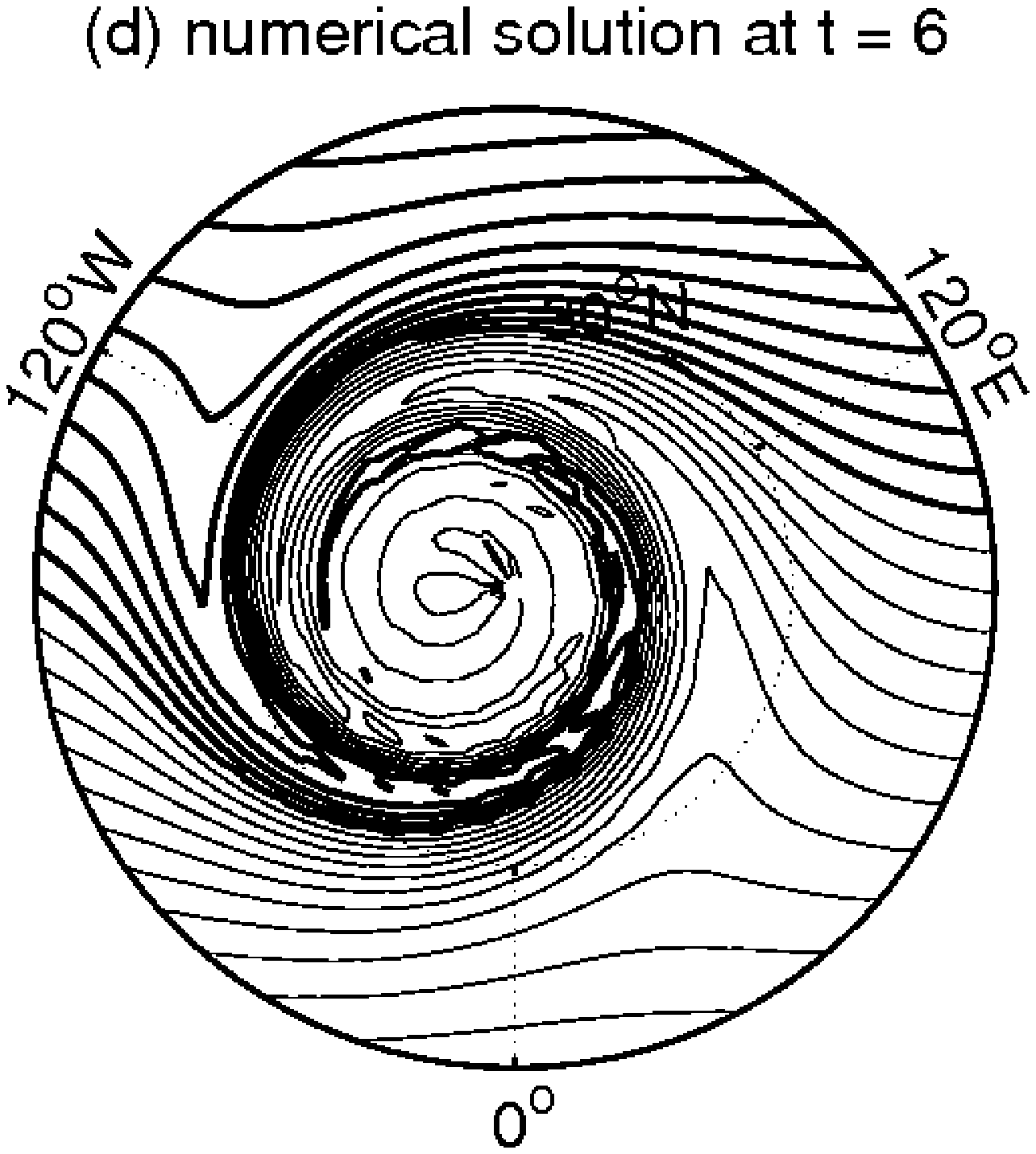}\end{center}

\caption{Results of a polar vortex simulation over the sphere. 
The exact solution and its numerical approximation at time 
$t=3$ can be found in panels (a) and (b), respectively. 
Contours plotted between 0.5 and 1.5 with contour interval 0.05. 
Panels (c) and (d) display the same results for time $t=6$.}
\label{figure3}
\end{figure}

Simulations are performed using a $128 \times 64$ grid and a step
size of $\Delta t = 0.05$. The filter (\ref{smoothing})
is not applied. The exact solution (evaluated over the given
grid) and its numerical approximation at times $t=3$ and $t=6$ are displayed 
in Fig.~\ref{figure3}. The relative $l_1$, $l_2$ and $l_\infty$ errors
(as defined in \cite{Zerroukat2}) can be found in Table \ref{table2}. 
These errors are comparable to the errors reported in 
\cite{Nair1,Zerroukat2} for the standard SL bicubic interpolation approach.

\bhead{Rotating shallow-water equations in planar geometry}

To demonstrate the behavior of the new advection scheme
under a time-dependent and compressible velocity field, 
we consider the shallow-water equations (SWEs) 
on an $f$-plane \cite{Durran98,salmon99}:
\begin{eqnarray} 
\frac{Du}{Dt} & = & +fv-g\mu _{x},\label{eq:unreg_u}\\
\frac{Dv}{Dt} & = & -fu-g\mu _{y},\label{eq:unreg_v}\\
\frac{D \mu }{Dt} & = & -\mu \,(u_{x}+v_{y}) .\label{eq:mu}
\end{eqnarray}
Here $\mu =\mu \left(x,y,t\right)$ is the fluid
depth, $g$ is the gravitational constant, and $f$
is twice the (constant) angular velocity of the reference plane.

Let $H$ denote the maximum value of $\mu$ over the whole fluid
domain. We also introduce the fluid depth perturbation
$\tilde \mu = \mu-H$. The perturbation satisfies the continuity equation
\begin{equation} \label{contpert}
\frac{D\tilde \mu}{Dt} = -\tilde \mu \,(u_x + v_y)
\end{equation}
which we solve numerically using the newly proposed scheme. The overall
time stepping procedure is given by the semi-Lagrangian St\"ormer-Verlet
(SLSV) method proposed by {\sc Reich} \cite{reich_bit06} with only equation
(5.7) from \cite{reich_bit06} being replaced by the following steps:
\begin{itemize}
\item[(i)] 
\[
\mu^{n+1/2-\varepsilon} = \mu^n - \frac{\Delta t H}{2}
\left[ u_x + v_y\right]^{n+1/2-\varepsilon}
\]
\item[(ii)] Solve (\ref{contpert}) over a full time step 
using the newly proposed scheme with velocities 
$(u^{n+1/2-\varepsilon},v^{n+1/2-\varepsilon})$ and initial
fluid depth perturbation $\tilde \mu^{n+1/2-\varepsilon} = 
\mu^{n+1/2-\varepsilon} - H$. Denote the resulting fluid
depth by $\mu^{n+1/2+\varepsilon} = \tilde \mu^{n+1/2+\varepsilon}+H$.
\item[(iii)]
\[
\mu^{n+1} = \mu^{n+1/2+\varepsilon} - \frac{\Delta t H}{2}
\left[ u_x + v_y\right]^{n+1/2+\varepsilon}
\]
\end{itemize}

The method has been implemented using the standard C-grid 
\cite{Durran98} over a 
double periodic domain with $L_x = L_y =$ 3840 km
(see \cite{ASL4} for details). The grid size is
$\Delta x = \Delta y = $ 60 km. The time step is $\Delta t = $ 20 min
and the value of $f$ corresponds to an $f$-plane at 45$^o$ latitude.
The reference height of the fluid is set to $H = $ 9665 m. The Rossby
radius of deformation is $L_R \approx $ 3000 km. Initial
conditions are chosen as in \cite{ASL4,reich_bit06} and results
are displayed in an identical format for direct comparison.

\begin{figure} 
\begin{center}$\qquad \quad $SLSV$\qquad \qquad \qquad \qquad \qquad \quad
 $SLSV-SISL\end{center}

\begin{center}\includegraphics[  scale=0.23]{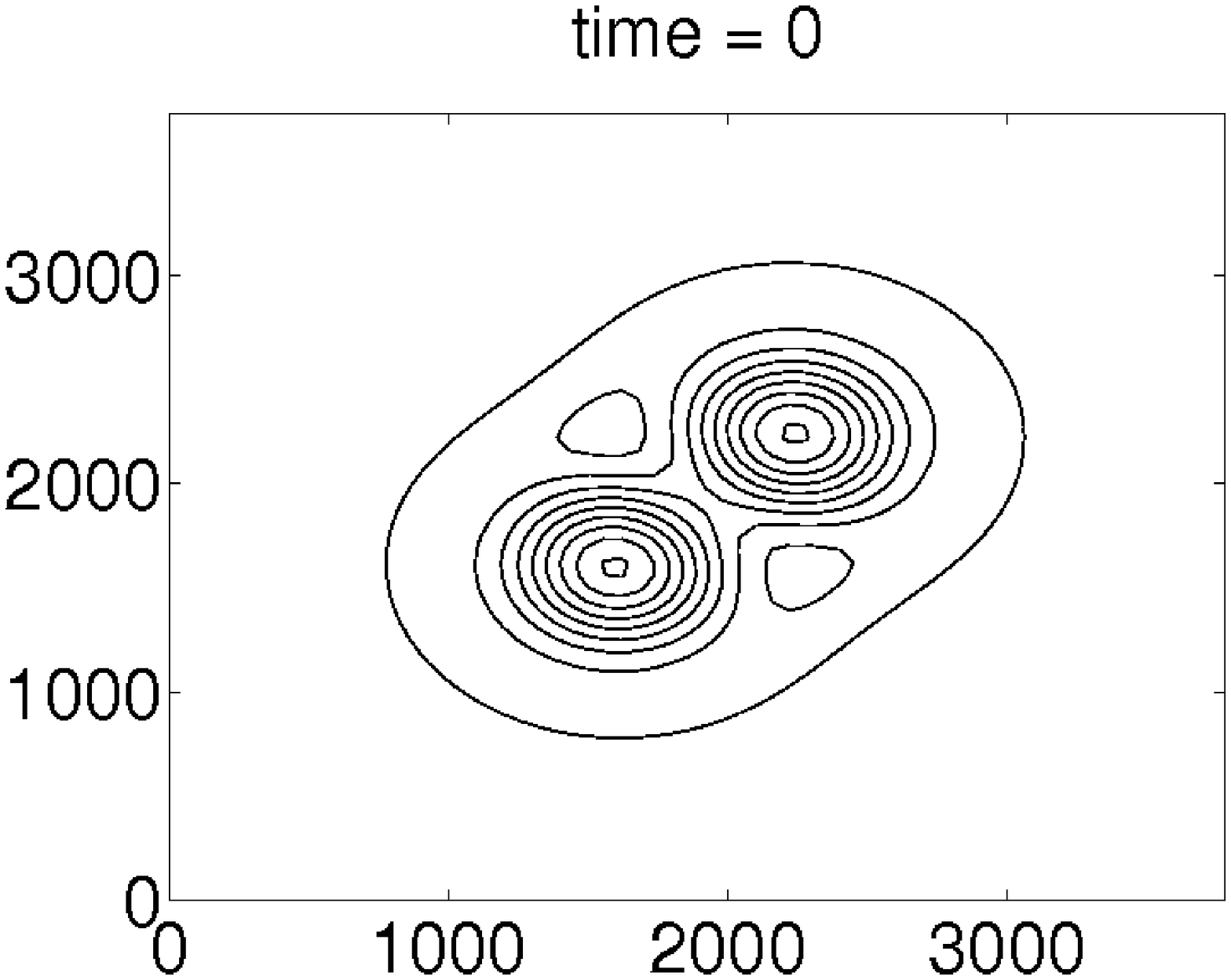}$\qquad$\includegraphics[  scale=0.23]{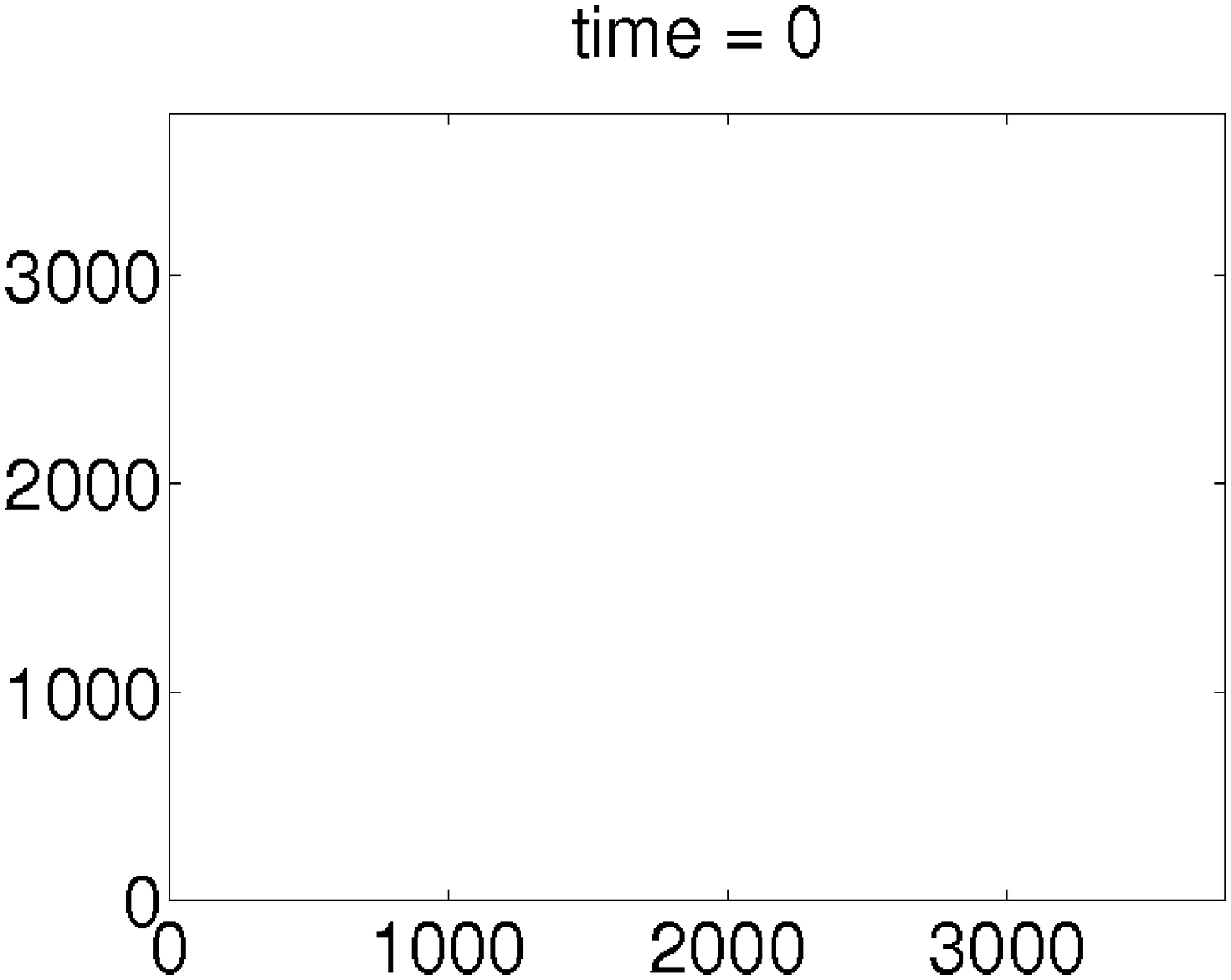}\end{center}

\begin{center}\includegraphics[  scale=0.23]{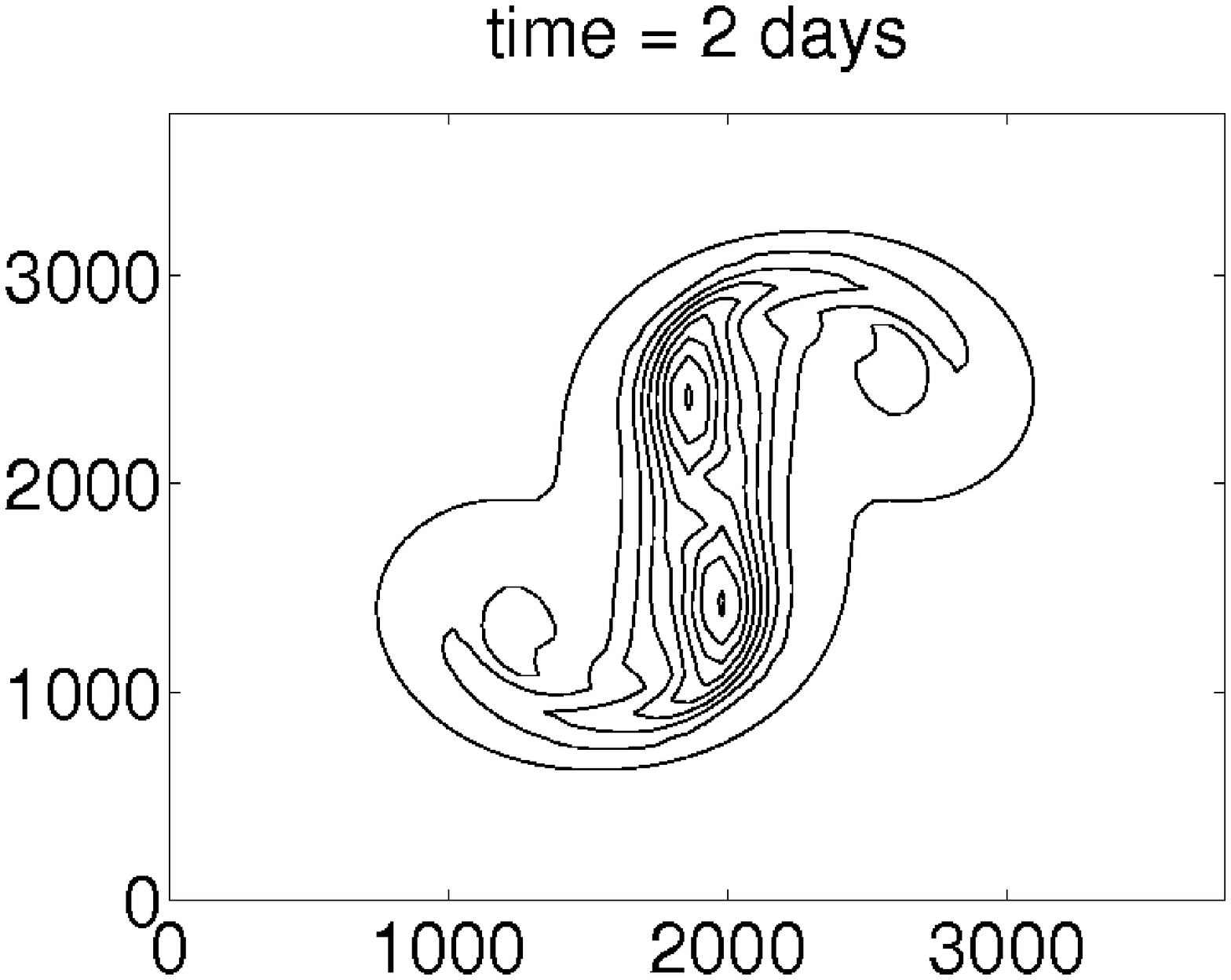}$\qquad$\includegraphics[  scale=0.23]{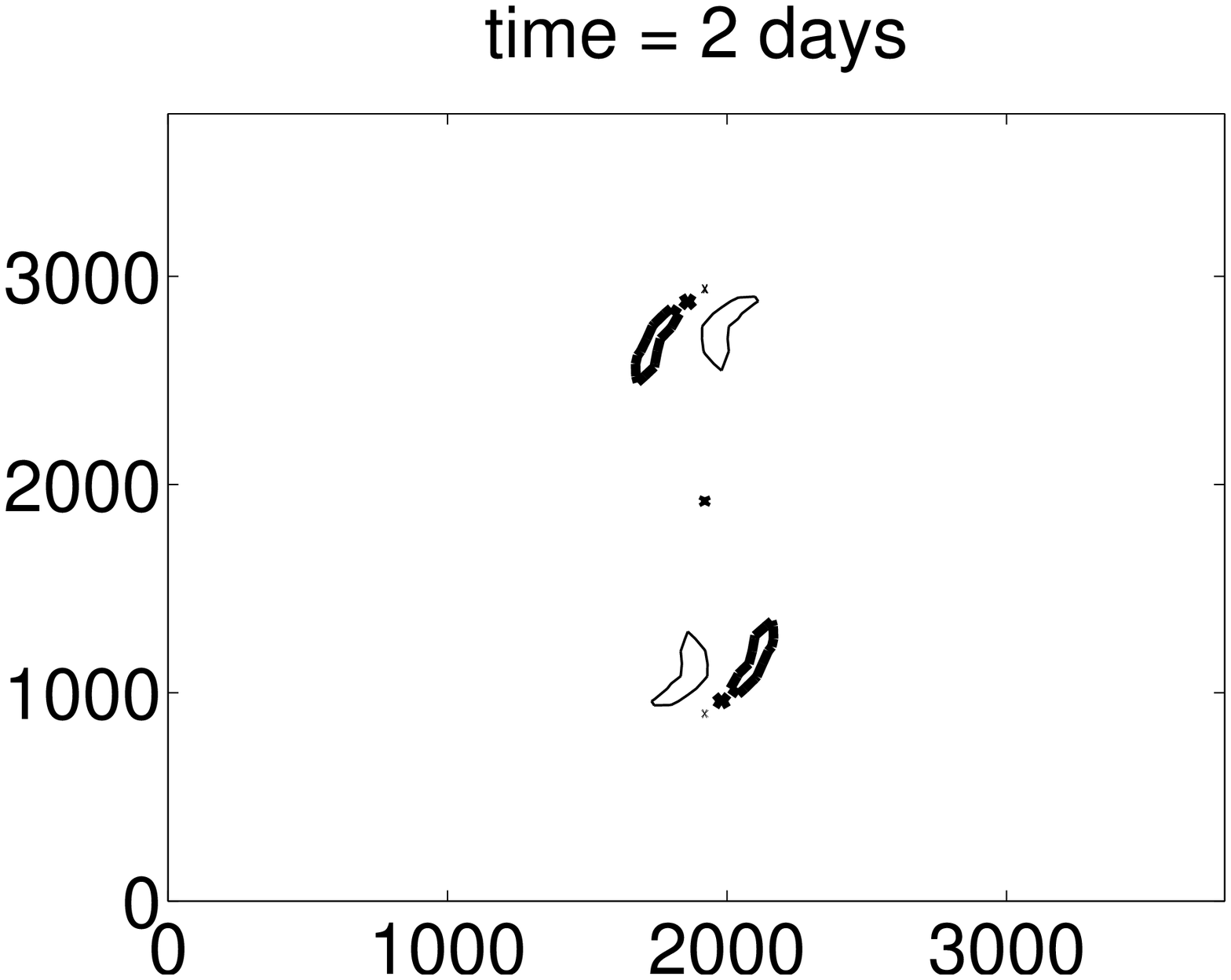}\end{center}

\begin{center}\includegraphics[  scale=0.23]{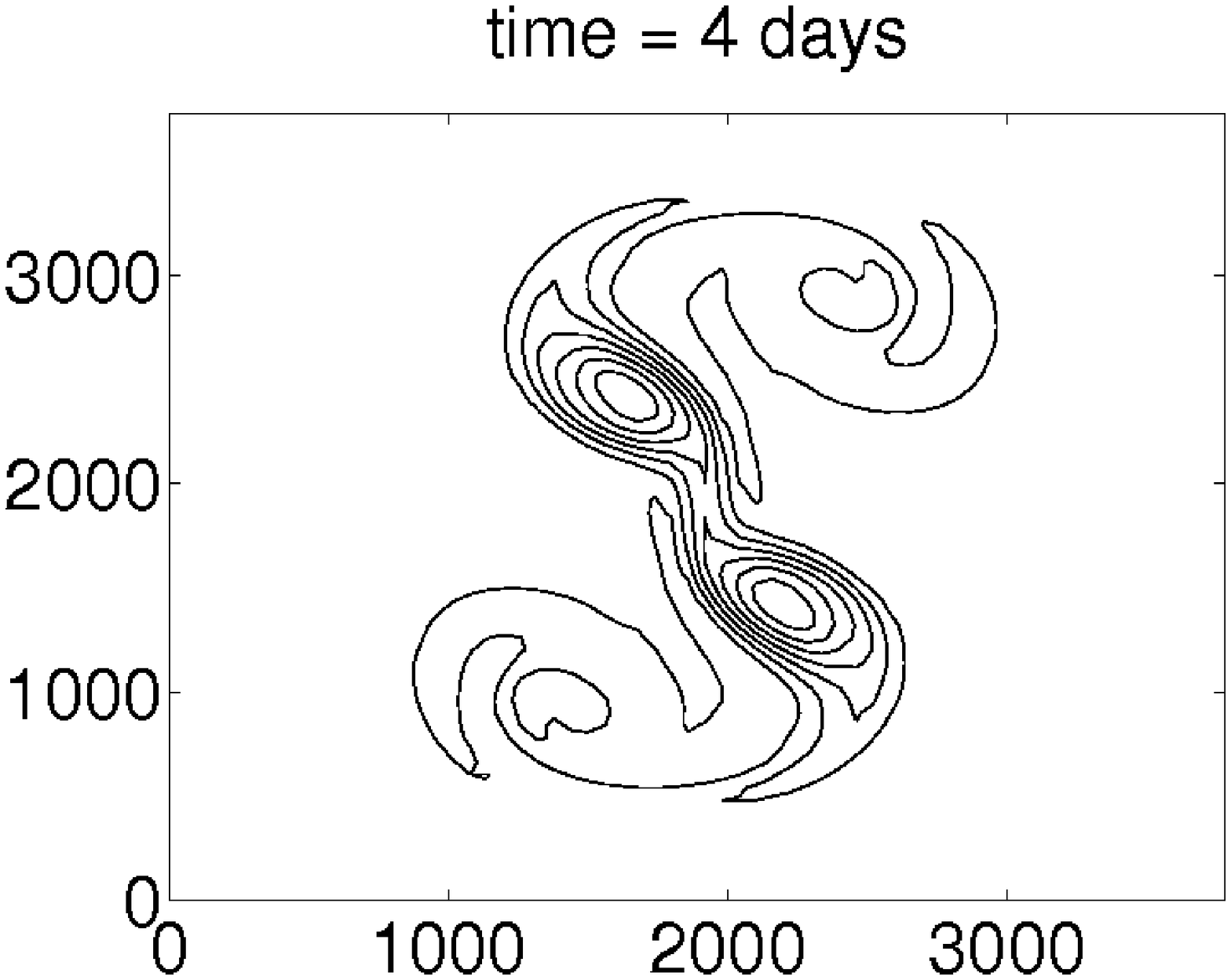}$\qquad$\includegraphics[  scale=0.23]{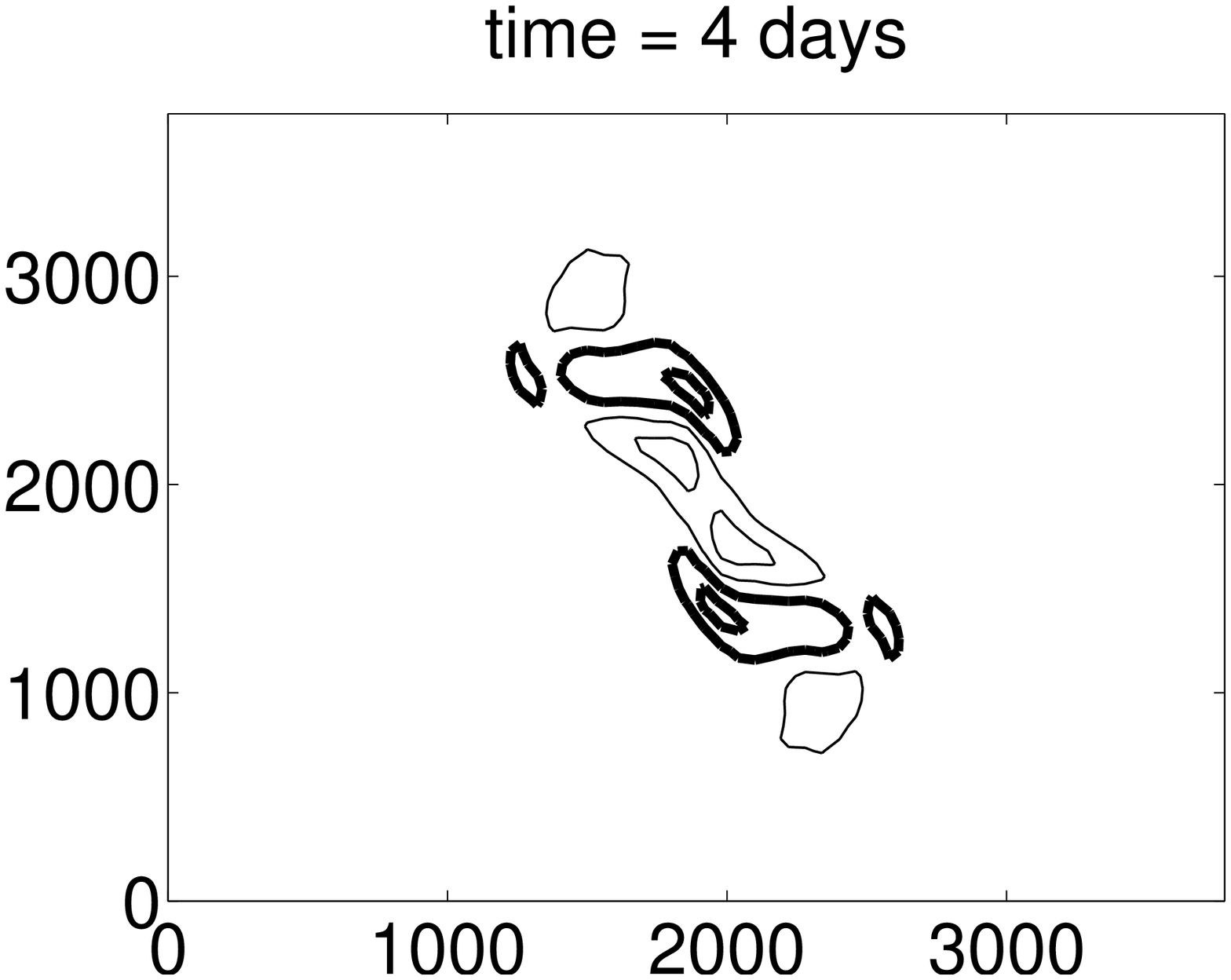}\end{center}

\begin{center}\includegraphics[  scale=0.23]{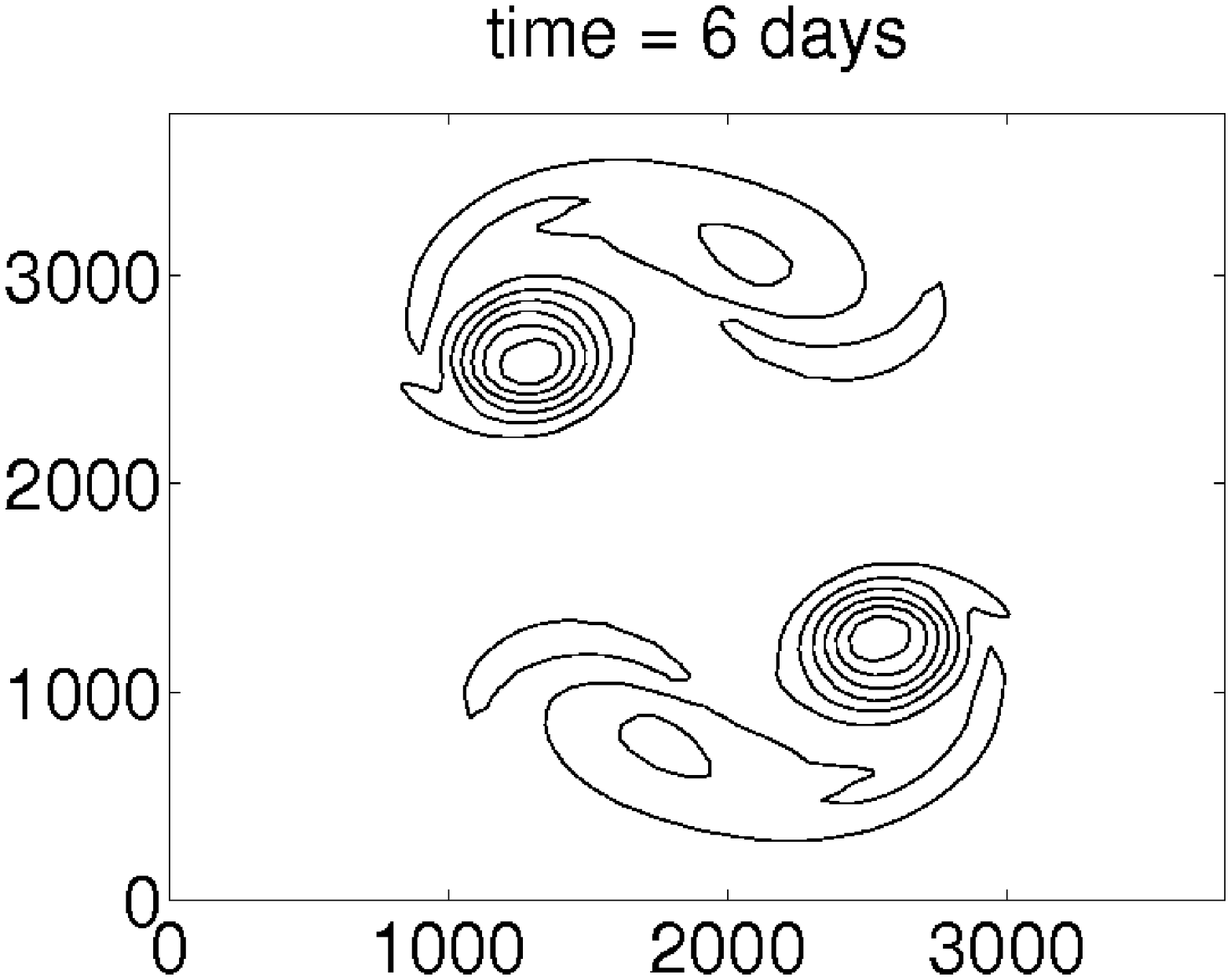}$\qquad$\includegraphics[  scale=0.23]{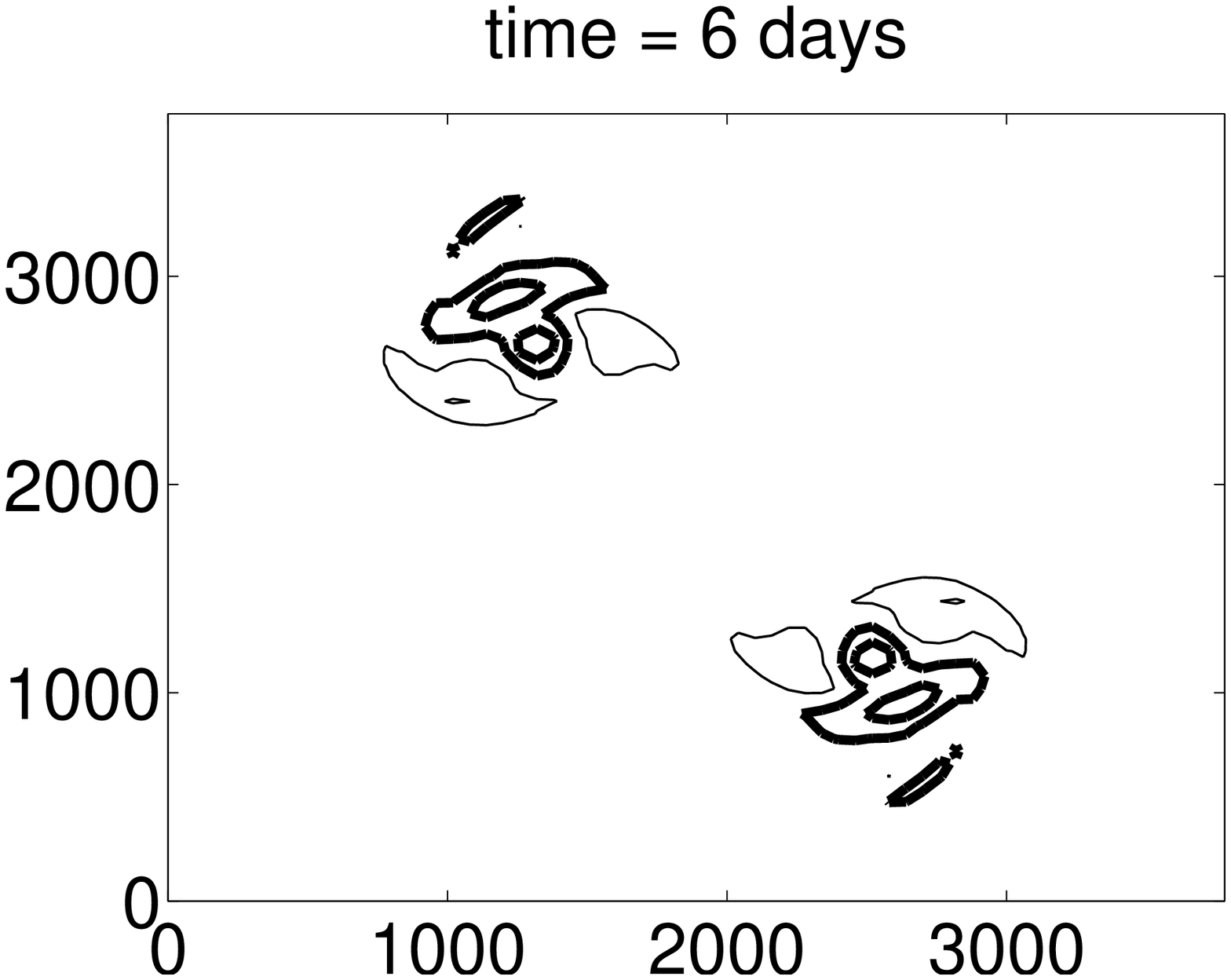}\end{center}

\caption{\label{cap:newfigure2} Left panels: 
Computed time evolution, from initial time
to $t=6$ days, of PV over the domain $(x,y)\in [0,3840\, \mathrm{km}]\times [0,3840\, \mathrm{km}]$
using the semi-Lagrangian St\"ormer-Verlet (SLSV) method 
with time step $\Delta t=20\, \textrm{min}$. Contours plotted 
between $6.4\times 10^{-8}\, \mathrm{m}^{-1}\mathrm{s}^{-1}$
and $2.2\times 10^{-7}\, \mathrm{m}^{-1}\mathrm{s}^{-1}$ with contour
interval $1.56\times 10^{-8}\, \mathrm{m}^{-1}\mathrm{s}^{-1}$. 
Right panels: Differences
(semi-Lagrangian St\"ormer-Verlet minus fully implicit semi-Lagrangian)
at corresponding times are plotted with a 10 times
smaller contour interval, where thin (thick) lines are positive (negative)
contours.}
\label{figure2}
\end{figure}

To assess the new discretization, results are compared to those from a
two-time-level semi-implicit semi-Lagrangian (SISL) method with a
standard bicubic interpolation approach to semi-Lagrangian advection
(see, e.g., \cite{McDonald,Temperton}). It is apparent from
Fig.~\ref{figure2} that both simulations yield similar results in
terms of potential vorticity advection.  Furthermore, the results
displayed in Fig.~\ref{figure2} are nearly identical to those
displayed in Fig.~6.1 of \cite{reich_bit06}.  The implication is that
the newly proposed advection scheme in manner very similar to the
traditional SL interpolation scheme for this particular test
problem. This result in not unexpected as the fluid depth remains
rather smooth throughout the simulation.

\ahead{Summary and outlook}
\label{summary}

A computationally efficient and mass conserving forward trajectory
semi-Lagrangian approach has been proposed for the solution of the
continuity equation (\ref{continuity}). At every time step a ``mass''
is assigned to each grid point which is then advected downstream
to a (Lagrangian) position. The gridded density at the next time step is
obtained by evaluating a bicubic spline representation with the advected
masses as weights. The main computational cost is given by the need to
invert tridiagonal linear systems in (\ref{system}). Computationally
efficient iterative or direct solvers are available. We also proposed an
extension of the advection scheme to spherical geometry. A further
generalization to 3D would be straightforward. Numerical experiments
show that the new advection scheme achieves accuracy comparable to 
standard non-concerving and published conserving SL schemes. 

We note that the proposed advection scheme can be used to advect
momenta according to
\begin{equation}
\frac{D}{Dt} (\rho {\bf u}) = -(\rho {\bf u} )\nabla \cdot {\bf u}.
\end{equation}
This possibility is particularly attractive in the context of the newly
proposed semi-Lagrangian St\"ormer-Verlet (SLSV) scheme \cite{reich_bit06}.

\acks
We would like to thank Nigel Wood for discussions and comments on
earlier drafts of this manuscript.

\nocite{*}
\bibliography{HPM_SemiLagrange}

\end{document}